\documentclass[%
 reprint,
superscriptaddress, 
nofootinbib, 
 amsmath,amssymb,
 aps,
]{revtex4-2}

\usepackage{graphicx}
\usepackage{dcolumn}
\usepackage{bm}
\usepackage[colorlinks, citecolor=blue]{hyperref}
\usepackage{longtable}
\usepackage{array,ulem}

\begin{document}

\title{A machine learning approach to tomographic pattern generation and classification of quantum states of light}

\author{Soumyabrata Paul}
\email{soumyabrata@physics.iitm.ac.in}
\affiliation{Department of Physics, Indian Institute of Technology Madras, Chennai 600036, India}
\affiliation{Center for Quantum Information, Communication and Computing (CQuICC), Indian Institute of Technology Madras, Chennai 600036, India}
\author{H. S. Subramania}
\affiliation{Telstra Limited, 242 Exhibition Street, Melbourne VIC 3000, Australia}
\author{S. Ramanan}
\affiliation{Department of Physics, Indian Institute of Technology Madras, Chennai 600036, India}
\affiliation{Center for Quantum Information, Communication and Computing (CQuICC), Indian Institute of Technology Madras, Chennai 600036, India}
\author{V. Balakrishnan}
\affiliation{Center for Quantum Information, Communication and Computing (CQuICC), Indian Institute of Technology Madras, Chennai 600036, India}
\author{S. Lakshmibala}
\affiliation{Center for Quantum Information, Communication and Computing (CQuICC), Indian Institute of Technology Madras, Chennai 600036, India}

\date{\today}

\begin{abstract}
Optical tomograms can be envisaged as patterns. The Wasserstein generative adversarial network (WGAN) algorithm provides a platform to train the machine to compare patterns corresponding to input and generated tomograms. Using a deep-learning framework with two convolutional neural networks and WGAN, we have trained the machine to generate tomograms of Fock states, coherent states (CS) and the single photon added CS ($1$-PACS). The training process was continued until the Wasserstein distance between the input and output tomographic patterns levelled off at a low value. The mean photon number,  variances and higher moments were extracted directly from the generated tomograms, to distinguish between different Fock states and also between the CS and the $1$-PACS, without using an additional classifier neural network. The robustness of our results has been verified using two error models and also with different colormaps that define the tomographic patterns. We have examined if the training program successfully reflected some of the findings in a recent experiment in which state reconstruction was carried out to establish that the fidelities between an amplified CS, an optimal CS and a $1$-PACS  were close to unity, over a range of parameter values. By training the machine to reproduce tomograms corresponding to these specific states, and comparing the mean photon numbers of these states obtained directly from the tomograms, we have established that the variations in these observables reflect the experimental trends. State reconstruction from tomograms could be challenging, in general, since the Hilbert space associated with quantized light is large. The tomographic approach provides a viable alternative to detailed state reconstruction. Our work demonstrates the use of machine learning to generate optical tomograms from which the states can be directly characterized. 
\end{abstract}
\maketitle

\section{Introduction\label{sec:introduction}}
\begin{figure}
\centering
\includegraphics[width=0.45\textwidth]{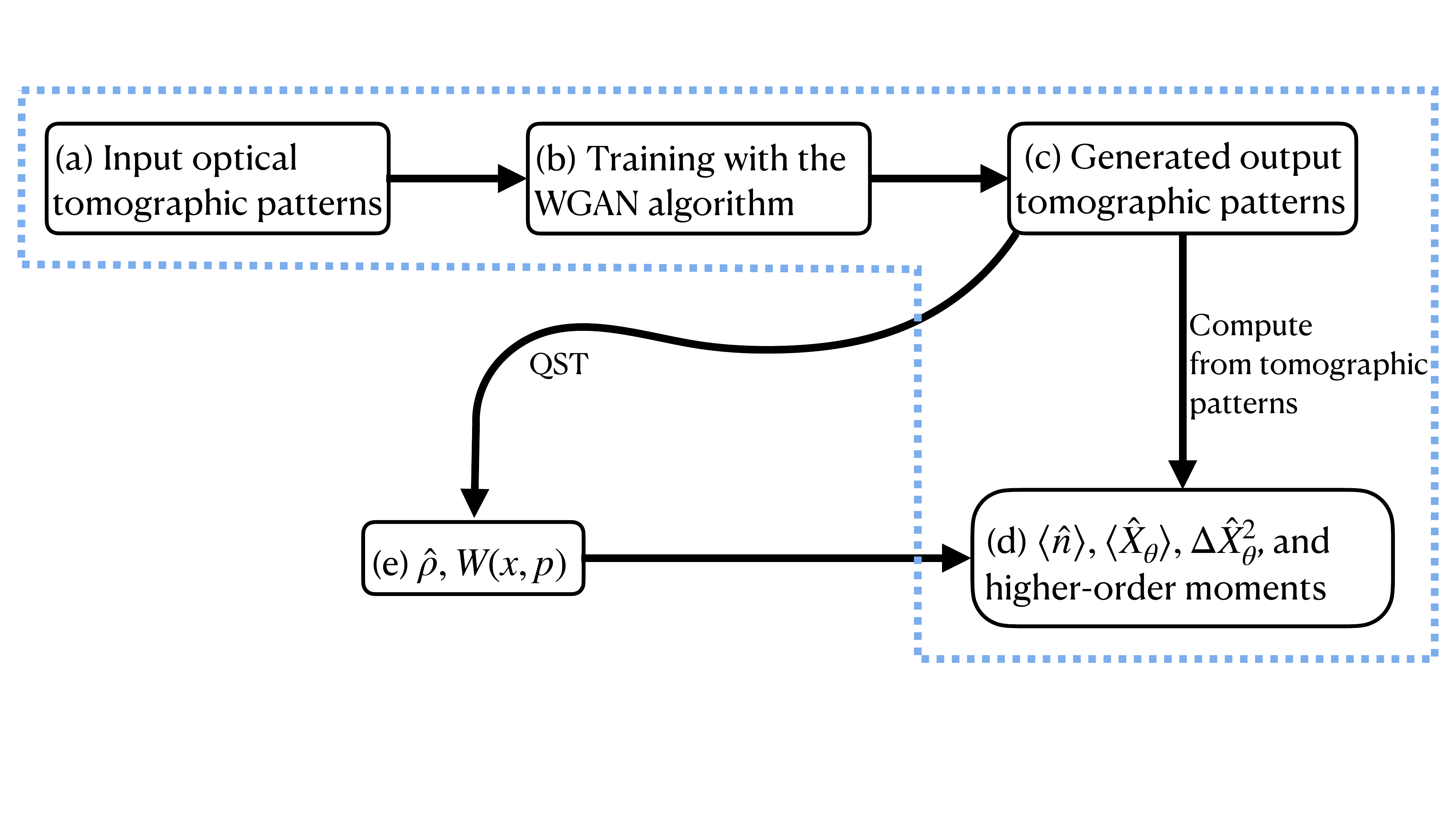}
\caption{{\bf Schematic of the procedure used for machine learning}. In the flowchart above, (a) the tomogram corresponding to a quantum state of light can be represented as a pattern. Here a collection of optical tomograms of each of three states, namely, the photon number states, coherent states, and $1$-photon added coherent states are separately used for pattern training. (b) The generator and the discriminator (also called a critic) are trained simultaneously with one of these three collections of tomograms. Training is through several convolutional layers. Wasserstein distance quantifies the extent of divergence between input and generated patterns. Feedback loop between discriminator and generator till discriminator acknowledges that generator has been reasonably trained. (c) After reasonably successful training, output tomographic patterns are available. Random samples from a large number of output tomograms used for step (d). (d) Directly compute with ease the mean photon number $\langle \hat{n} \rangle$ and quadrature variance $\Delta {\hat X}_{\theta}^{2}$, corresponding to quadrature angle $\theta$, from each output tomogram. Compare with corresponding quantities computed from input patterns to estimate extent of errors. (e) Quantum state tomography (QST) to obtain the state $\hat{\rho}$ and the Wigner function $W(x, p)$ from output tomograms, (for characterizing the state) is circumvented here. QST could pose challenges for continuous variable systems. Further, from the quantities $\langle \hat{n} \rangle$, $\Delta {\hat X}_{\theta}^{2}$ and higher-order moments computed readily from output tomographic patterns, classification is straightforward without an additional neural network. Our approach is contained in the blue dashed box, and provides a viable alternative to the QST procedure.}
\label{fig:schematic_flowchart}
\end{figure}
In recent years, deep neural network-based machine learning tools have found tremendous application in diverse areas of quantum physics, such as state tomography~\cite{Carrasquilla:2019, Torlai:2019, Palmieri:2020, Tiunov:2020, Ahmed:2021a, Ahmed:2021b, Ghosh:2021, Danaci:2021, Hsieh:2022, Cha:2022, Ma:2024, Hsieh:2024, Ma:2025, Wu:2025}, designing new experiments~\cite{Krenn:2016, Molesky:2018, Nichols:2019, Krenn:2021, RuizGonzalez:2023, Krenn:2025, Landgraf:2025}, feedback control~\cite{Borah:2021, Sivak:2022, Reuer:2023, Porotti:2023, Vaidhyanathan:2024, Hutin:2025}, and error correction~\cite{Fosel:2018, Valenti:2019, Sweke:2021, Olle:2024, Zen:2024, Puviani:2025}. Detailed reviews of some of these aspects are available, for instance in~\cite{Melko:2019, Krenn:2023, Wang:2023, Gebhart:2023, Acampora:2025, Dawid:2025}. In the context of quantum optics with which we will be concerned, substantial work has been carried out on various aspects of quantum state tomography (QST) which is the procedure for reconstructing an unknown state of light from experimental data. Typically, the data is a collection of probability density functions (PDFs) of the field quadratures corresponding to different angles in a homodyne setup. Each angle $\theta$ defines a tomographic slice (projection), and each PDF is a function of $\theta$ and the corresponding quadrature variable $X_{\theta}$. A quorum of these PDFs is the optical tomogram. Reconstruction of the quantum state from the optical tomogram requires, in principle, an infinite number of such PDFs~\cite{Ibort:2009}. Since this is not feasible in practice, it is important to find out the optimal number of PDFs and the corresponding quadrature angles for a reasonable reconstruction of the Wigner function or the density matrix corresponding to a state. It has been pointed out in~\cite{Leonhardt:1996, Gandhari:2024} that in the case of single mode states of light, a finite number of equispaced tomographic slices suffices for a reasonable reconstruction of the density matrix from the tomogram. 
  
Quantifiers such as the fidelity between states can be computed only from the corresponding density matrices. While state reconstruction is straightforward for single-mode examples, the bipartite cases prove to be challenging. For instance, even in the Jaynes-Cummings model of a two-level atom interacting with the radiation field, the field state at different instants during temporal evolution of the system was reconstructed only as recently as 2017~\cite{Lv:2017}. It is known that for multimode systems the task is extremely challenging in general. Techniques such as maximum likelihood estimation (MLE) have been used in QST (see, for instance,~\cite{Hradil:1997}) to determine the density matrix that is most likely to have produced the observed experimental data. These techniques are, in general, computationally demanding. 

Alternatively machine learning (ML) programs have been used for state reconstruction. These have primarily focused on training the machine to identify and reproduce known states of light (density matrices) which were used as inputs for training. These density matrices were generated at the end of the training session, with high fidelity, using fewer data points and iterative steps than both accelerated projected-gradient-based procedures and iterative maximum-likelihood estimation, using a conditional generative adversarial network (CGAN)~\cite{Ahmed:2021a, Ahmed:2021b}. However, the challenges faced in the context of multimode systems remain. 

In contrast, in the case of a wide range of bipartite systems, investigations have been carried out to determine the optimal number of slices required to identify entanglement indicators directly from the tomograms~\cite{Sharmila:2019, Sharmila:2020, Sharmila:2022} and also to reasonably reproduce sum uncertainty relations~\cite{Soumyabrata:2023}. 

Further, the tomographic approach described earlier has been used in ML to identify bipartite entanglement. Bipartite tomograms corresponding to a random set of input density matrices with stellar rank $\leqslant 2$, were used as input in this ML program to show that with merely four bipartite tomographic slices, entanglement can be identified~\cite{Gao:2024}. In spirit, this is analogous to medical imaging, where $3$D structures are reconstructed from intensity profiles obtained in an optimal number of $2$D slices. 

Since the tomographic approach avoids the challenges of QST, and can be successfully used to characterize single-mode states and estimate bipartite entanglement, it is worthwhile to exploit the ML approach in the context of tomograms themselves. Our primary motivation is two-fold: (a) to train the machine to generate tomograms which mimic the input tomograms successfully, using an appropriate algorithm, and (b) to compute properties of states corresponding to these tomograms directly, from a limited number of tomographic slices themselves, without using further ML protocols or auxiliary networks. In the following sections, we will characterize different states by computing the mean photon number, the variances and higher moments, from the generated tomograms. In Fig.~\ref{fig:schematic_flowchart} we present a flowchart indicating the tomographic approach (within the blue dotted line) and where QST differs from it. 
 
As illustrated in the next section, the optical tomogram corresponding to a state can be visualized as a pattern. Since the Wasserstein distance is known to be useful in pattern comparison and recognition, we have used the Wasserstein generative adversarial network (WGAN) algorithm for our purpose, with tomograms both as inputs and generated outputs. In fact it has been demonstrated in the context of Fock states, the coherent state, the squeezed vacuum state, and the even coherent state, that in certain contexts of relevance to experiments, the Wasserstein distance is a good indicator of the extent of addition of photons to an appropriately chosen reference state. Whereas the fidelity between two Fock states, and between states obtained by adding different even numbers of photons to the squeezed vacuum (or the even coherent state) vanishes, the corresponding Wasserstein distances are nonzero and provide clear indicators to distinguish between the states~\cite{Soumyabrata:2025a, Soumyabrata:2025b, Soumyabrata:2025c}. Thus, the WGAN is an appropriate choice for the ML program at hand. It distinguishes between the true and generated tomographic probability distributions, and can be used to monitor the quality of the generated state through different epochs. Further, unlike other generic GANs, it provides a smooth, non-vanishing gradient even when the distributions do not overlap, leading to a reliable and continuous signal for the generator to learn. After the ML program, we have also extracted the mean photon number, the quadrature variances and the higher moments directly from the generated output tomograms, and assessed the extent of errors, without using an additional neural network. 

The three classes of input states that we have considered are the following: (i) the coherent state (CS) $|\alpha\rangle$ which is the closest analog to a classical radiation field (with mean photon number $|\alpha|^{2}$ and positive Wigner function), (ii) the Fock state $|n\rangle$ which has $n$ photons and negativity in its Wigner function, and (iii) the single-photon added CS (or $1$-PACS) $|\alpha, 1\rangle$, with mean photon number $2L_{2}(-|\alpha|^{2}) / L_{1}(-|\alpha|^{2}) - 1$, where $L_{k}$ is the Laguerre polynomial of order $k$, and with negative regions in the Wigner function~\cite{Agarwal:1991}. Since $|\alpha,1\rangle$ interpolates between $|\alpha\rangle$ and $|1\rangle$, it provides an ideal route to also examine the classical-to-quantum transition~\cite{Zavatta:2004a, Zavatta:2005}. 

This paper is organized as follows. In Secs.~\ref{sec:optical_tomograms} and~\ref{sec:wgan_network_arch_dataset_prep} we briefly review the salient features of an optical tomogram, the WGAN algorithm, and the convolutional neural network-based architecture that we have used for our purpose. In Sec.~\ref{sec:results_Fock_CS_PACS_Fadrny_experiment} we present our results on the efficacy of the WGAN algorithm in training the machine to recognize input tomograms for the Fock states, the CS and the $1$-PACS. In all three cases, the plots of the mean photon numbers and the variances in the $x$-quadrature computed from the generated tomograms have been presented and discussed. A detailed investigation on the role played by the choice of colormaps has been included. We have also compared other quantities obtained from these generated tomograms with those from a recent experiment reported in the literature~\cite{Fadrny:2024}, to establish the extent of success of the training program. We conclude with some remarks and outlook in Sec.~\ref{sec:concluding_remarks}. Appendix~\ref{sec:appendix_numerical_arrays} extends our analysis to input and output data represented as numerical arrays, rather than the RGB format used in the main text. In Appendix~\ref{sec:appendix_choice_of_err_tol}, we elaborate on the error tolerance required for correctly classifying the states in this work. Appendix~\ref{sec:appendix_noise_models} comprises plots corresponding to noise models used in the training process, and Appendix~\ref{sec:appendix_real_expts} considers real experimental data for training the WGAN algorithm. In Appendix~\ref{sec:appendix_Wasserstein_based_confusion_matrix}, we quantify the performance of our classification procedure using a Wasserstein distance-based confusion matrix. Finally, plots corresponding to the $p$-quadrature are presented in Appendix~\ref{sec:appendix_variances_p_quadratures}. 

\section{Input optical tomograms for training\label{sec:optical_tomograms}}
\begin{figure}[h]
\centering
\includegraphics[width=0.45\textwidth]{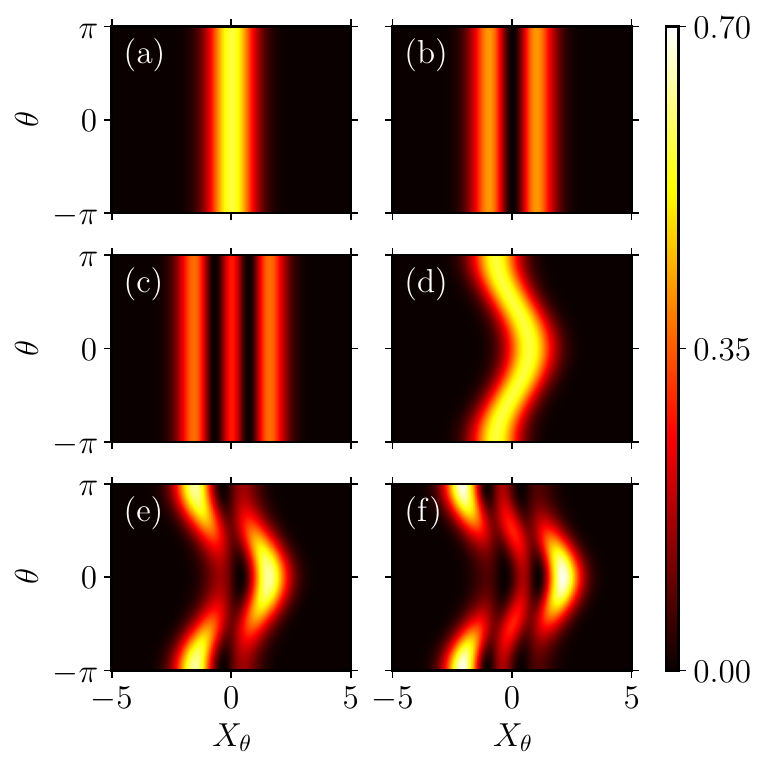}
\caption{Optical tomograms $w(X_{\theta}, \theta)$ corresponding to six single-mode states of the radiation field. Here, (a), (b) and (c) are the tomograms for the $0$, $1$ and $2$-photon states respectively, and (d), (e) and (f) correspond to a CS $|\alpha\rangle$, the $1$-photon added CS $|\alpha,1\rangle$ and the $2$-photon added CS $|\alpha,2\rangle$, with $\alpha=0.5$. For this value of $\alpha$ and the range of values selected for the machine learning procedure, the probability distribution functions and hence the tomograms, are predominantly in the range $X_{\theta} \in [-5, 5]$. We note that the single vertical cut in (b) and (e) corresponds to one photon, and two vertical cuts in (c) and (f) to two photons. The number of zeroes of the PDFs in the different quadratures and their location determine the number and location of the cuts. }
\label{fig:fig_tomogram_Fock_0_1_2_PACS_alpha_0.5_m_0_1_2_panel}
\end{figure}
Given a single-mode radiation field with photon creation and annihilation operators ($\hat{a}^{\dagger}, \hat{a}$), we consider the set of rotated quadrature operators~\cite{Ibort:2009},
\begin{equation}
\hat{\mathbb{X}}_{\theta} = \left( \hat{a}^{\dagger} e^{i\theta} + \hat{a} e^{-i\theta} \right) / \sqrt{2},
\label{eq:qudrature_ops}
\end{equation}
where $\theta$ ($0 \leqslant \theta < \pi$), is the phase of the local oscillator in the standard homodyne measurement setup. $\theta = 0$ and $\pi/2$ correspond to the $x$ and $p$ quadratures respectively. The set $\{\hat{\mathbb{X}}_{\theta}\}$ constitutes a quorum of observables that carry complete information about a given state of light with density operator ${\hat{\rho}}$. The optical tomogram is given by~\cite{Lvovsky:2009} 
\begin{equation} 
w(X_{\theta}, \theta) = \langle X_{\theta}, \theta | \hat{\rho} | \ X_{\theta}, \theta \rangle,
\label{eq:w_singlemode}
\end{equation}
where $\hat{\mathbb{X}}_{\theta} |X_{\theta}, \theta \rangle = X_{\theta}|X_{\theta}, \theta \rangle$. We note that $\{|X_{\theta}, \theta \rangle\}$ forms a complete basis for every $\theta$. We denote the basis corresponding to $\theta=0$ by $|X\rangle$. For a normalized pure state, $\hat{\rho}=|\psi\rangle\langle\psi|$ and $w(X_{\theta}, \theta)=|\psi(X_{\theta}, \theta)|^{2}$ for a given $\theta$. Every tomogram satisfies the completeness relation
\begin{equation} 
\int_{-\infty}^{\infty} {d}X_{\theta}~w(X_{\theta}, \theta) = 1 ~ \forall ~ \theta,
\label{eq:w_singlemode_completeness}
\end{equation}
and the symmetry property
\begin{equation} 
w(X_{\theta}, \theta + \pi) = w(-X_{\theta}, \theta).
\label{eq:w_singlemode_symmetry}
\end{equation}
Each tomogram is a collection of histograms corresponding to the quadrature operators, and is visualized with $X_{\theta}$ as the abscissa, and $\theta$ as the ordinate. For state reconstruction, although it is sufficient to work with the range $0 \leqslant \theta < \pi$, the tomogram plotted for $0 \leqslant \theta < 2\pi$ (equivalently, $-\pi \leqslant \theta \leqslant \pi$ as in Fig.~\ref{fig:fig_tomogram_Fock_0_1_2_PACS_alpha_0.5_m_0_1_2_panel}) helps visualize various features better. It is evident that as a consequence of Eq.~\ref{eq:w_singlemode_symmetry} tomograms plotted from $0 \leqslant \theta \leqslant 2\pi$ will be reflected about the vertical axis when compared with their corresponding counterparts plotted from $-\pi \leqslant \theta \leqslant \pi$. 

The tomogram comprises only the diagonal elements of $\hat{\rho}$. These are readily available from experimental measurements. For each $\theta$ we have a tomographic slice with an associated probability density $|\psi(X_{\theta}, \theta)|^{2}$. Our aim is to train the machine to learn to identify a range of tomograms (constructed with an optimal number of finite slices) corresponding to important single-mode states of light such as the photon number states, the coherent state and the $1$-photon added coherent state.

\section{WGAN, network architecture and dataset preparation\label{sec:wgan_network_arch_dataset_prep}}
The WGAN algorithm involves two neural networks (NNs), namely, the generator $G$ and the discriminator $D$. The two network architectures are presented in Tables~\ref{tab:generator_params} and~\ref{tab:discriminator_params} respectively. $G$ generates data (which, to begin with, will not in general be close to the real data). $D$ distinguishes between the real and the generated data. The input data (tomograms) are fed in batches to the discriminator network. The two networks are trained iteratively until $G$ becomes good at generating data very close to the real data. The loss function (the difference between the two datasets) is defined using the Kantorovich-Rubinstein duality~\cite{Arjovsky:2017, Gulrajani:2017}, and is quantified by the Wasserstein distance~\footnote{Equation~(\ref{eq:Kantorovich_Rubinstein_duality}) for the Wasserstein distance between two datasets is a generalization of the Wasserstein distance (or $1$-Wasserstein distance $W_{1}$) between two normalized PDFs $f(x)$ and $g(x)$ with corresponding cumulative distribution functions $F(x)$ and $G(x)$ respectively. The latter is given by $W_{1}(F, G) = {\int}_{-\infty}^{\infty} dx \left| F(x) - G(x) \right|$.\label{fn:wass_dist_bw_pdf}}
\begin{equation}
W(\mathbb{P}_{r}, \mathbb{P}_{g}) = \underset{G}{{\rm min}} ~ \underset{D \in \mathcal{D}}{{\rm max}} ~ \Big( \underset{\pmb{x} \sim \mathbb{P}_{r}}{\mathbb{E}}[D(\pmb{x})] - \underset{\pmb{\tilde{x}} \sim \mathbb{P}_{g}}{\mathbb{E}}[D(\pmb{\tilde{x}})] \Big).
\label{eq:Kantorovich_Rubinstein_duality}
\end{equation}
Here, $\pmb{x}$ and $\pmb{\tilde{x}}$ are the real and generated samples respectively. $\mathbb{E}$ and $\mathcal{D}$ denote the expectation value computed over batches, and the family of $1$-Lipschitz functions, respectively. ($1$-Lipschitz constrained NNs are amenable to ML tasks, due to their improved stability and better defense against adversarial attacks~\cite{Bethune:2022}.) To impose the $1$-Lipschitz condition on the discriminator $D$, we used the method of gradient penalty~\cite{Gulrajani:2017}. $\mathbb{P}_{r}$ ($\mathbb{P}_{g}$) is the real (generated) data distribution. It is important to note that the loss function for the discriminator comprises both the terms in the RHS of Eq.~(\ref{eq:Kantorovich_Rubinstein_duality}), whereas the loss for the generator arises only from the second term in the RHS of Eq.~(\ref{eq:Kantorovich_Rubinstein_duality}). 

\begin{figure}
\centering
\includegraphics[width=0.45\textwidth]{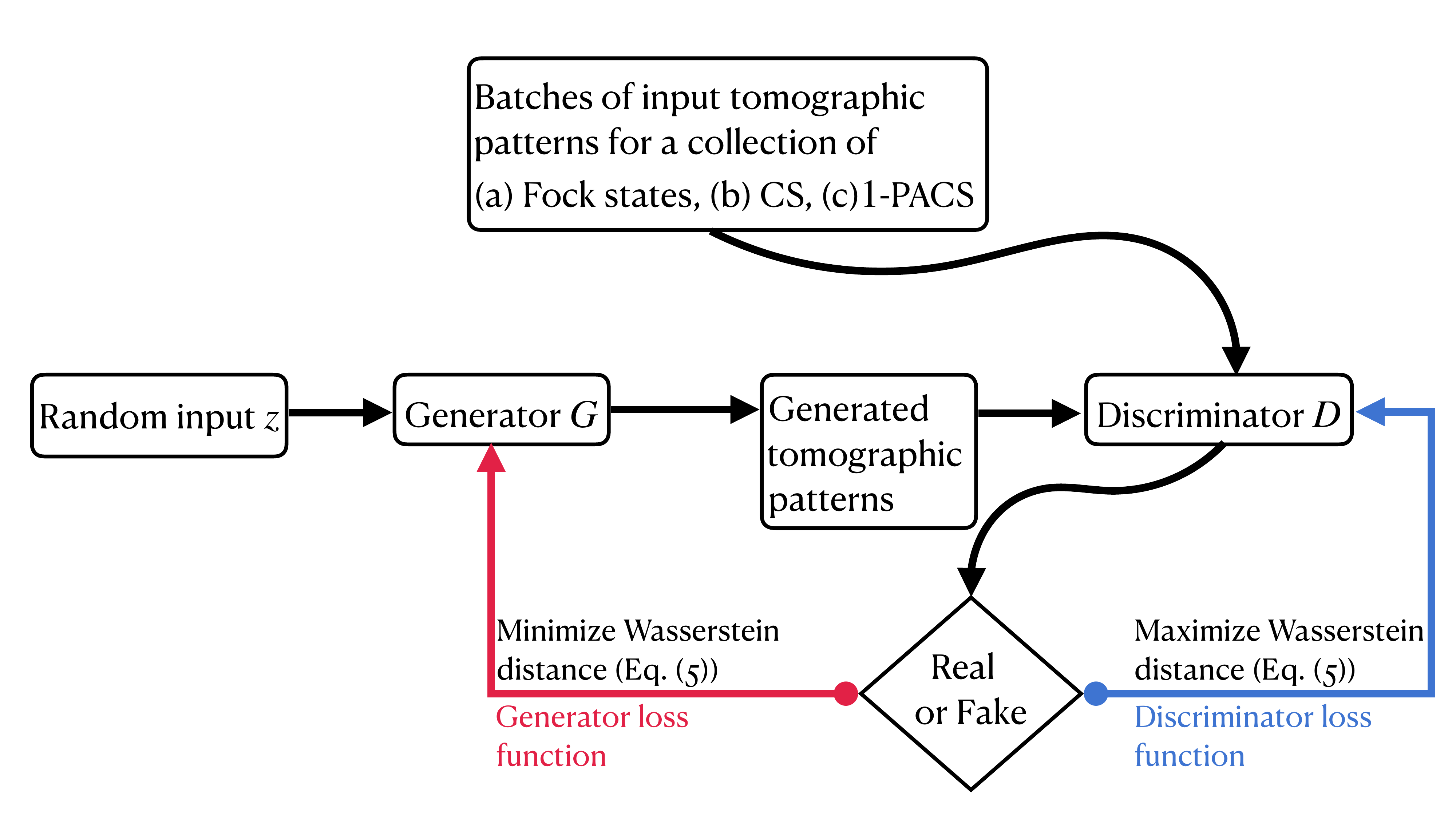}
\caption{{\bf Training with the Wasserstein generative adversarial network (WGAN) algorithm: Schematic of the WGAN framework}. The WGAN algorithm is composed of two deep neural networks, a generator $G$ and a discriminator $D$. ($G$ and $D$ are two separate convolutional neural networks (CNNs) in this work.) $G$ is initialized with a random input $z$ which is sampled from a uniform/ Gaussian distribution. $G$ learns the `real' (input training) tomograms corresponding to (a) Fock states, (b) CS and (c) $1$-PACS, through feedback from $D$. $D$ on the other hand, compares the output from $G$ (which are the `fake' tomograms) and the real tomograms, and gives feedback to $G$. (The feedbacks are backpropagations for adjusting weights.) The discriminator's (generator's) objective is to maximize (minimize) the Wasserstein distance (Eq.~(\ref{eq:Kantorovich_Rubinstein_duality})). The learning process is complete when Eq.~(\ref{eq:Kantorovich_Rubinstein_duality}) has saturated (equivalently when $G$ generates samples mimicking the real input tomograms). Other architectural details of the CNNs and the values of the hyperparameters used are given in Tables~\ref{tab:generator_params} and~\ref{tab:discriminator_params}, and Sec.~\ref{sec:wgan_network_arch_dataset_prep}, respectively.}
\label{fig:WGAN_schematic}
\end{figure}

Each training loop (also referred to, as an epoch) involves training the discriminator five times and the generator once. Every image/tomogram in the training dataset is subject to one training loop. In what follows, we have trained the machine using the WGAN algorithm for $25000$ epochs for each class of states, and for each error threshold that we set initially. A schematic of the WGAN framework is given in Fig.~\ref{fig:WGAN_schematic}. 

We used the Adam optimizer, a popular gradient-based optimization algorithm in deep learning, known for its adaptive learning rate and moment-based updates. In particular, this optimizer combines the advantages of two extensions of stochastic gradient descent, namely, AdaGrad and RMSProp, by maintaining running averages of the moments. We set the decay rate for the moment $\beta_{1}$ to be $0$, the decay rate for squared gradients $\beta_{2}$ to be $0.9$, and the learning rate to be $0.0001$. These values correspond to the configuration in~\cite{Gulrajani:2017}. This choice ensures stable convergence, especially in training adversarial networks where gradient dynamics can be volatile. 

Further, we had set the batch size to be $64$. (The batch size refers to the number of training samples processed together in one forward and backward pass.) We note that both the stability and efficiency of training are sensitive to the batch size. Smaller batches offer noisier gradient estimates but faster updates, and larger batches provide smoother gradients at the cost of memory and computation. 

\begin{table}[h!]
\centering
\begin{tabular}{c|c|c}
\hline \hline
Layer & Output shape & No. of parameters \\
\hline
Conv2D-T(4, 1, 0) & (512, 4, 4) & 819 200 \\
Batch normalization & (512, 4, 4) & 1024 \\
Conv2D-T(4, 2, 1) & (256, 8, 8) & 2 097 152 \\
Batch normalization & (256, 8, 8) & 512 \\
Conv2D-T(4, 2, 1) & (128, 16, 16) & 524 288 \\
Batch normalization & (128, 16, 16) & 256 \\
Conv2D-T(4, 2, 1) & (64, 32, 23) & 131 072 \\
Batch normalization & (64, 32, 23) & 128 \\
Conv2D-T(4, 2, 1) & (32, 64, 64) & 32 768 \\
Batch normalization & (32, 64, 64) & 64 \\
Conv2D-T(4, 2, 1) & (3, 128, 128) & 1539 \\
Total parameters & & 3 608 003 \\
\hline \hline
\end{tabular}
\caption{Layers, shapes and trainable parameters of the Generator NN: The tranpose convolution layers are denoted by Conv2D-T($k$, $s$, $p$) where $k$, $s$ and $p$ are the kernel size, stride and padding respectively. Batch normalization is done between the first five Conv2D-T layers. After the final (sixth) Conv2D-T layer, the activation function $\tanh$ is used. The resulting output serves as an input to the Discriminator NN for training purposes.}
\label{tab:generator_params}
\end{table}

\begin{table}[h!]
\centering
\begin{tabular}{c|c|c}
\hline \hline
Layer & Output shape & No. of parameters \\
\hline
Conv2D(4, 2, 1) & (16, 64, 64) & 784 \\
Conv2D(4, 2, 1) & (32, 32, 32) & 8192 \\
Instance normalization & (32, 32, 32) & 64 \\
Conv2D(4, 2, 1) & (64, 16, 16) & 32 768 \\
Instance normalization & (64, 16, 16) & 128 \\
Conv2D(4, 2, 1) & (128, 8, 8) & 131 072 \\
Instance normalization & (128, 8, 8) & 256 \\
Conv2D(4, 2, 1) & (256, 4, 4) & 524 288 \\
Instance normalization & (256, 4, 4) & 512 \\
Conv2D(4, 2, 0) & (1, 1, 1) & 4097 \\
Total parameters & & 702 161 \\
\hline \hline
\end{tabular}
\caption{Layers, shapes and trainable parameters of the Discriminator NN: The convolution layers are denoted by Conv2D($k$, $s$, $p$) where $k$, $s$ and $p$ are the kernel size, stride and padding respectively. After the first Conv2D layer and after each instance normalization layer, the activation function LeakyReLU is applied. The output after the final (sixth) Conv2D layer is used to calculate the Wasserstein loss function given by Eq.~(\ref{eq:Kantorovich_Rubinstein_duality}).}
\label{tab:discriminator_params}
\end{table}

For every single-mode optical tomogram that we considered, $\theta \in [-\pi, \pi]$, and for each $\theta$, $X_{\theta} \in [-5, 5]$ as shown in Fig.~\ref{fig:fig_tomogram_Fock_0_1_2_PACS_alpha_0.5_m_0_1_2_panel}. Both these ranges were divided into $128$ equal partitions. Each tomogram used for training (input tomogram) can be visualized as an image, obtained using an appropriate colormap. The details of the colormaps that we have used are given in subsequent sections. After including the $3$ RGB pixel values of the image, the grid dimension was $3 \times 128 \times 128$. 

An alternative procedure that can be used is by working with numerical arrays (raw tomographic data), both as inputs and outputs. However, as demonstrated in detail in Appendix~\ref{sec:appendix_numerical_arrays}, the results obtained in this case are only comparable to and not significantly better than those obtained using images, i.e., RGB data as both input to, and output from the network. In a subsequent section we will point out the specific limitation that is seen while working with images or with numerical arrays. For our purpose therefore, we will carry out the machine learning program with images, without any loss of generality. 

Since we have been working with known states for demonstration purposes, the tomograms were computed by numerically obtaining the diagonal elements of the density matrix in different quadratures using Eq.~(\ref{eq:w_singlemode}). It is often advantageous computationally to expand $w(X_{\theta}, \theta)$ in the Fock basis, for pure states~\cite{Filippov:2011}. (In the case of unknown states the diagonal elements of the density matrix would be obtained from the homodyne setup.) We note that no information about the off-diagonal elements of the density matrix is required in this tomographic approach. As mentioned in the Introduction, our aim is not merely to train the machine to produce visually good quality tomographic patterns, but also to calculate experimentally relevant quantities such as the mean photon number, squeezing properties of the state, quadrature variances and all higher moments of the field quadratures from the generated tomograms themselves. This will demonstrate the efficiency of the ML protocol. 

The input states that we have considered are (a) Fock states $|n\rangle$ ($n = 0,1,2\dots,5$), (b) the CS and (c) the $1$-PACS $|\alpha,1\rangle$, with $\alpha = 0.0, \sqrt{0.1}, \sqrt{0.3}, \sqrt{0.5}, 1.0$. Sixteen identical copies of the tomogram corresponding to each state (i.e., for a given value of $n$ or $\alpha$) are used as the training dataset for the WGAN algorithm. The training program uses either a set of Fock states with different values of $n$, or a set of CS or $1$-PACS with different values of $\alpha$. We have found that $25000$ epochs are sufficient for training purposes. 

Further, to theoretically address experimental imperfections due to detector inefficiencies, measurement errors and systematic errors during state preparation and the homodyne procedure, two different error models were considered. In (a) $w(X_{\theta}, \theta)$ was transformed to $[1 + \mathcal{U}(\epsilon)] w(X_{\theta}, \theta)$, and in (b) to $[1 + \mathcal{N}(\mu=0, \sigma^{2}=\epsilon^{2})] w(X_{\theta}, \theta)$, corresponding to a random set $\{(X_{\theta}, \theta)\}$ as described below. Here $\mathcal{U}(\epsilon)$ and $\mathcal{N}(\mu, \sigma^{2})$ represent points sampled from the uniform distribution and the normal distribution (with mean $\mu$ and variance $\sigma^{2}$), respectively. Two maximum error thresholds were considered by setting $\epsilon = 0.10$ (marginal error) and $0.25$ (realistic error in experiments using homodyne procedures, without accounting for losses in the fiber~\cite{Fadrny:2024}). For every state the training data had four sets of tomograms, each set comprising four identical copies. One set of tomograms were error free, and the other three sets had $2.5\%$, $5\%$ and $7.5\%$, respectively of randomly chosen $(X_{\theta}, \theta)$ points, where the errors models were applied. Therefore, the training dataset comprised sixteen tomograms which included noisy tomograms for every state, and a particular choice of the error model. The plots of the mean photon number and $\Delta \hat{x}^{2}$ corresponding to Fock states are presented in Appendix~\ref{sec:appendix_noise_models}, with the given threshold values of $\epsilon$. It is evident from these plots that noise addition with the above mentioned threshold values of $\epsilon$ do not bring about significant changes in our conclusions. 

We have also verified the efficacy of the tomographic approach in the context of two real experiments that were performed to produce and characterize single-photon Fock states~\cite{Hsieh:2024, Zavatta:2004b}. The states were produced using a parametric amplifier with overall efficiency $\eta = 0.631$ and $\eta = 0.574$, respectively. The experimentally produced state corresponds to a density matrix $(1 - \eta)|0\rangle \langle0| + \eta|1\rangle \langle1|$. The experimenters have also reported the best fit PDF given by 
\begin{equation}
P(X;\eta) = \sqrt{\frac{2}{\pi}}\bigl[1 - \eta(1 - 4X^2)\bigr] e^{-2X^2}. 
\label{eq:expt_Fock_1_PDF_model}
\end{equation} 
(We note that Eq.~(\ref{eq:expt_Fock_1_PDF_model}) is independent of the choice of the quadrature because it is a single-photon state.) Hence, we now have access to the full tomogram obtained from the experiments. We have therefore used this as the input for the WGAN algorithm with the two different values of $\eta$, namely, $0.574$ and $0.631$. We have established that distinguishing between states is clearly possible in the experimental context as well. The corresponding results are summarized in Appendix~\ref{sec:appendix_real_expts}. 

\section{Generation of tomographic slices with WGAN,  and characterization\label{sec:results_Fock_CS_PACS_Fadrny_experiment}}
\begin{figure}
\centering
\includegraphics[width=0.45\textwidth]{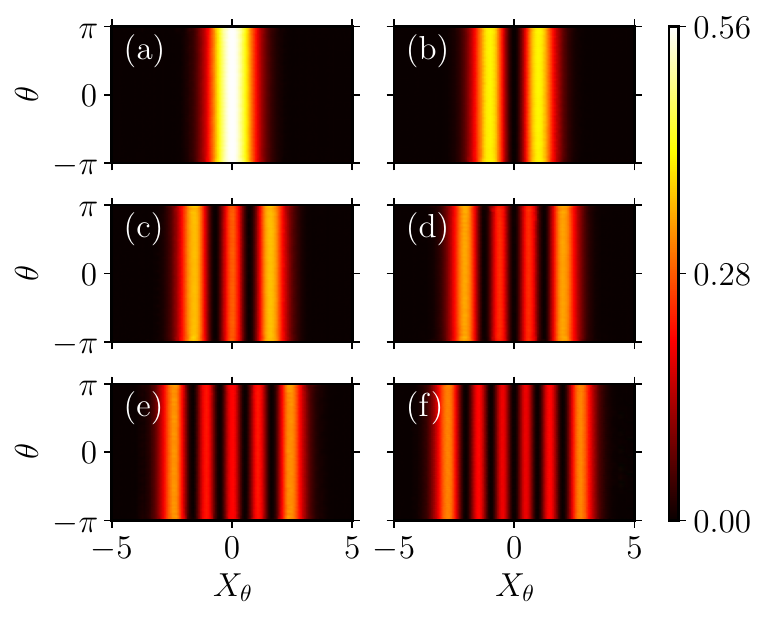}
\caption{Representative generated samples for the Fock states $|n\rangle$ obtained after training with the WGAN algorithm for $25000$ epochs. Here $n=0,1,2,3,4$ and $5$ in (a), (b), (c), (d), (e) and (f), respectively. It is evident that the intensity distribution does not change with $\theta$ in these tomograms. This feature is discussed further in Sec.~\ref{sec:Fock}.}
\label{fig:generated_Fock_tomograms_no_errors_panel}
\end{figure}
The output tomograms are characterized by computing the mean photon number $\langle {\hat a}^{\dagger}{\hat a} \rangle$ (denoted by $\langle \hat{n} \rangle$), and also the quadrature variances $\Delta \hat{X}^{2}_{\theta}$ (for a range of values of $\theta$) corresponding to each generated sample. We recall that for $\theta=0$ and $\pi/2$ we obtain the variances along the $x$ and $p$ quadratures, respectively. In each of the three cases, namely, the Fock states, the CS and the $1$-PACS, we have considered $500$ output samples of the generated tomograms. (We have also verified that by increasing the number of generated samples to $2000$, there are no significant changes in our results.) The WGAN algorithm randomly generates tomograms with different values of $n$ for Fock states, and $\alpha$ for the CS and the $1$-PACS. Representative generated samples for the three classes of states are shown in Figs.~\ref{fig:generated_Fock_tomograms_no_errors_panel},~\ref{fig:generated_CS_tomograms_no_errors_panel} and~\ref{fig:generated_PACS_tomograms_no_errors_panel}. To calculate $\langle \hat{n} \rangle$ we used the expression~\cite{Wunsche:1996}
\begin{align}
\langle {\hat a}^{\dagger m}{\hat a}^{n} \rangle &= C_{mn} \sum_{k=0}^{m+n} \exp \Big\{ - \frac{ik(m-n)\pi}{m+n+1} \Big\} \times\nonumber \\
& \int_{-\infty}^{\infty} dX_{\theta}~w \Big( X_{\theta}, \frac{k\pi}{m+n+1} \Big) H_{m+n}(X_{\theta}),
\label{eq:mean_ph_num_Wunsche}
\end{align}
where $C_{mn}=(m!n!)/\big\{ (m+n+1)!2^{(m+n)/2} \big\}$. Setting $m=n=1$ it follows from Eq.~(\ref{eq:mean_ph_num_Wunsche}) that computation of the mean photon number of a generic state involves the PDFs corresponding to $\theta=0, \pi/3$ and $2\pi/3$.

The computation of the quadrature variances from the tomograms can be readily carried out. For $\theta=0$, $|\psi(x)|^{2}$ can be extracted directly from the tomogram, and hence any expectation value of the form $\langle{\hat x}^{k}\rangle = \int_{-\infty}^{\infty} dx~|\psi(x)|^{2} x^{k}$ ($k>0$) can be calculated. A similar exercise can be carried out for computing the variance and higher moments in any other quadrature. We have evaluated $\Delta \hat{X}^{2}_{\theta}$ in different quadratures for every sample that we have generated. Representative plots are presented in the subsequent sections for $\theta=0$ ($x$-quadrature) and $\pi/2$ ($p$-quadrature) alone.

\subsection{Photon number states\label{sec:Fock}}
\begin{figure}
\centering
\includegraphics[width=0.45\textwidth]{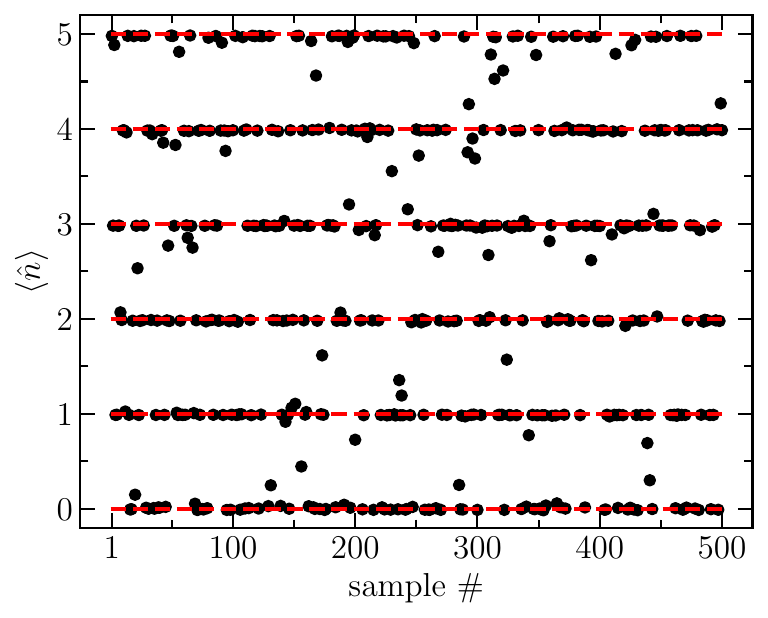}
\caption{The mean photon number $\langle \hat{n} \rangle$ (black circles) for the Fock states $|n\rangle~(n = 0, 1, \dots, 5)$ computed from $500$ randomly chosen generated tomograms. $\langle \hat{n} \rangle$ has been calculated from the reconstructed PDFs along $\theta=0, \pi/3$ and $2\pi/3$, by setting $m=n=1$ in Eq.~(\ref{eq:mean_ph_num_Wunsche}). The red dashed lines are the theoretical values of $\langle \hat{n} \rangle$ corresponding to the Fock states considered. The computed value of $\langle \hat{n} \rangle$ is within $4\%$ error tolerance of the theoretical value for each generated state whose tomogram can be identified to clearly correspond to a particular photon number. Spurious samples (typically black circles midway between two consecutive horizontal red lines) do not fall under this category. These arise due to inherent issues in the generation process.} 
\label{fig:Fock_gen_samples_mean_photon_number_no_errors}
\end{figure}
The training program was carried out with a set of photon number states $|n\rangle~(n=0,1,2\dots,5)$ with $16$ identical copies of each Fock state. To begin with, the training was carried out without introducing errors in the input tomograms. In this case $500$ generated (output) tomograms were randomly selected after training, and the mean photon number and the quadrature variances were computed for each tomogram. 

Representative output tomograms are shown in Fig.~\ref{fig:generated_Fock_tomograms_no_errors_panel}. These tomograms reveal that for a given Fock state the intensity distribution is $\theta$ independent. This can be seen as follows. Consider the density matrix $\hat{\rho}=|n\rangle \langle n|$ corresponding to the Fock state $|n\rangle$. Using Eq.~(\ref{eq:w_singlemode}), the relation $|X_{\theta},\theta \rangle = e^{i\theta\hat{a}^{\dagger}\hat{a}}|X\rangle$, and that $|n\rangle$ is an eigenstate of $\hat{a}^{\dagger}\hat{a}$ it is straightforward to see that $w(X_{\theta}, \theta) = |\langle X|n \rangle|^{2} = |\psi_{n}(X)|^{2}$. Therefore for the photon number states, the corresponding optical tomograms are quadrature independent.

\begin{figure}
\centering
\includegraphics[width=0.45\textwidth]{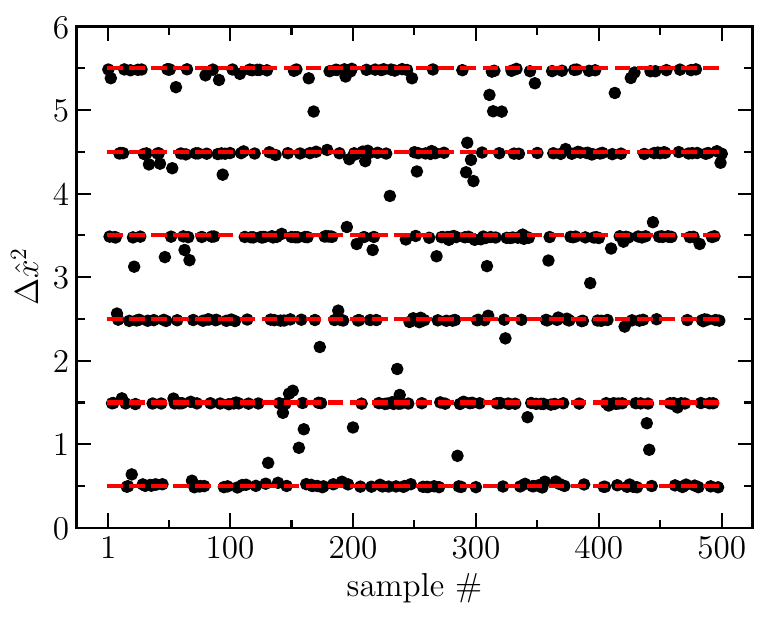}
\caption{The quadrature variance $\Delta \hat{x}^{2}$ (black circles) versus the sample number for $|n\rangle~(n=0,1,\dots,5)$ corresponding to $500$ randomly chosen generated samples, along the $x$-quadrature. This is evaluated using the reconstructed PDFs for $\theta=0$. The horizontal red dashed lines are the theoretical values of $\Delta \hat{x}^{2}$ corresponding to the Fock states considered. Most of the generated samples are within $4\%$ error tolerance of the theoretical values. Spurious samples arise due to inherent issues in the generation process.}
\label{fig:variance_gen_image_Fock_500_samples_theta_0.00000_no_errors}
\end{figure}
The expected (theoretically calculated) values of the mean photon number are given by the horizontal red dashed lines in Fig.~\ref{fig:Fock_gen_samples_mean_photon_number_no_errors}. Each black circle in Fig.~\ref{fig:Fock_gen_samples_mean_photon_number_no_errors} corresponds to $\langle \hat{n} \rangle$ computed from an output sample using Eq.~(\ref{eq:mean_ph_num_Wunsche}). The computed mean photon numbers are within $4\%$ error tolerance of the theoretical values.

A similar exercise has been carried out for a set of quadrature variances $\Delta \hat{X}^{2}_{\theta}$. The quadratures correspond to $\theta=0, \pi/4, \pi/3, \pi/2, 2\pi/3$ and $3\pi/4$, for each sample. As in the case of the mean photon number, barring spurious samples, the variances were seen to be within $4\%$ tolerance of the theoretical values. Here, we have only presented the variances in canonically conjugate slices ($\theta=0$ and $\pi/2$) versus the sample number in Figs.~\ref{fig:variance_gen_image_Fock_500_samples_theta_0.00000_no_errors} and~\ref{fig:variance_gen_image_Fock_500_samples_theta_1.57080_no_errors}, respectively. The quadrature variances corresponding to $|0\rangle$ both in the $x$ and the $p$ quadratures are expected to be ideally equal to $0.5$. From Figs.~\ref{fig:variance_gen_image_Fock_500_samples_theta_0.00000_no_errors} and~\ref{fig:variance_gen_image_Fock_500_samples_theta_1.57080_no_errors} it is evident that this is indeed true for most of the generated samples. The Heisenberg equality is satisfied in the case of most samples to within $4\%$ error tolerance and in a few cases up to $8\%$ error tolerance.

As mentioned in Sec.~\ref{sec:wgan_network_arch_dataset_prep}, error models have been studied in detail with Fock states. Plots of the mean photon number and $\Delta \hat{x}^{2}$  (Figs.~\ref{fig:Fock_gen_samples_mean_photon_number_error_model_b_eps_0.25} and~\ref{fig:variance_gen_image_Fock_500_samples_theta_0.00000_error_model_b_eps_0.25} respectively), obtained with noisy Fock states, trained with error model (b) and the threshold value of $\epsilon$ set to $0.25$, are presented in Appendix~\ref{sec:appendix_noise_models}. Comparing these plots with Figs.~\ref{fig:Fock_gen_samples_mean_photon_number_no_errors} and~\ref{fig:variance_gen_image_Fock_500_samples_theta_0.00000_no_errors} respectively, it is evident that the error models do not significantly change the conclusions drawn without introducing errors.

The genesis of the generated spurious states is both due to the WGAN algorithm and due to the choice of the colormap. While the former cannot be rectified since it is inherent to the generation process, we have pointed out (Sec.~\ref{sec:nonlinear_cmap_CS}) the manner in which issues arising due to the coarseness of the colormap can be reduced. For this purpose we have considered the PDFs corresponding to generated coherent states, whose characterization is carried out in the next section. 

\subsection{Coherent states\label{sec:CS}}
\begin{figure}
\centering
\includegraphics[width=0.45\textwidth]{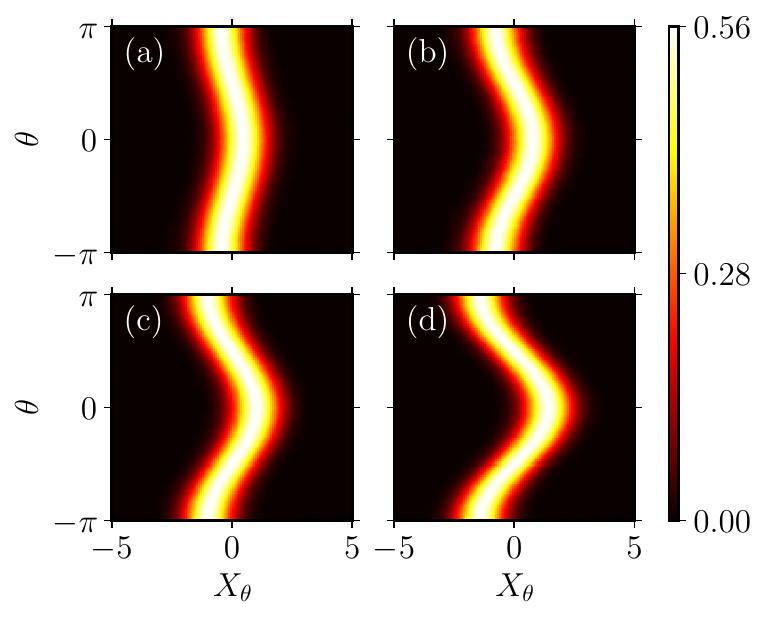}
\caption{Representative generated samples for the coherent states $|\alpha\rangle$ obtained after training with the WGAN algorithm for $25000$ epochs. Here $\alpha = \sqrt{0.1}, \sqrt{0.3}, \sqrt{0.5}$ and $1$ in (a), (b), (c) and (d), respectively.}
\label{fig:generated_CS_tomograms_no_errors_panel}
\end{figure}
\begin{figure}
\centering
\includegraphics[width=0.45\textwidth]{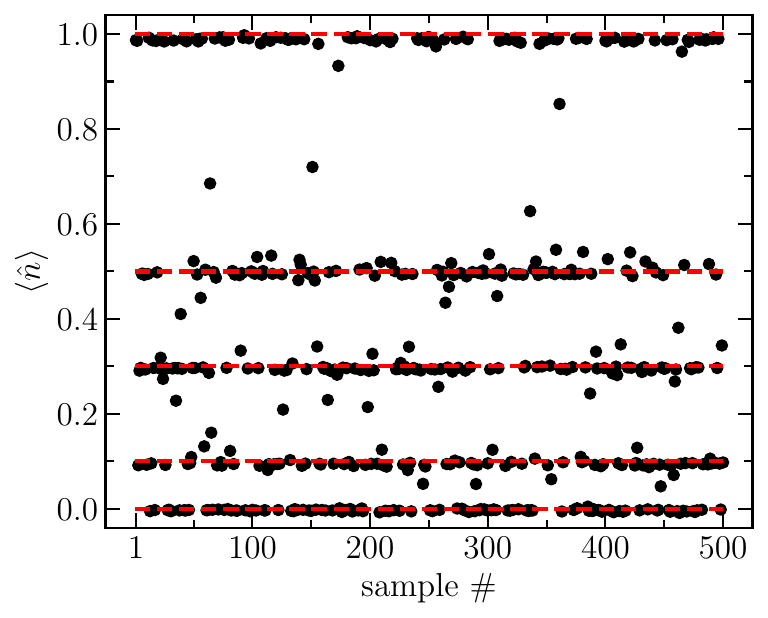}
\caption{The mean photon number $\langle \hat{n} \rangle$ (black circles) for the CS $|\alpha\rangle$ corresponding to each of the generated samples. This is evaluated using the reconstructed PDFs along $\theta=0, \pi/3$ and $2\pi/3$, and the expression in Eq.~(\ref{eq:mean_ph_num_Wunsche}) setting $m=n=1$. The horizontal red dashed lines represent the theoretical values of $\langle \hat{n} \rangle$ (i.e., $\alpha^{2}$) given by $0, 0.1, 0.3, 0.5$, and $1$. The computed value of $\langle \hat{n} \rangle$ is within $4\%$ of the theoretical value for most generated samples. As in the case of Fock states, spurious samples are outside this error tolerance and are inherent to the generation process.}
\label{fig:CS_gen_samples_mean_photon_number_no_errors}
\end{figure}
As in the case of Fock states, $500$ output tomograms of the CS were generated using the WGAN algorithm. To begin with, no errors were introduced in the input tomograms used for training. The corresponding representative output tomograms are shown in Fig.~\ref{fig:generated_CS_tomograms_no_errors_panel}. We note that in contrast to the tomograms for the Fock states, the intensity distribution in tomograms corresponding to the CS, depends on $\theta$. This can be traced back to the fact that the maximum of the corresponding PDFs (displaced Gaussians) varies with $\theta$. Input tomograms that were used for the training comprised five different values of $\alpha$, namely $0$, $\sqrt{0.1}$, $\sqrt{0.3}$, $\sqrt{0.5}$ and $1$. Each `good' output tomogram is expected to correspond to one of these values. Since these are coherent states, the mean photon number for good generated samples must be approximately $\alpha^{2}$ (indicated by the horizontal red dashed lines in Fig.~\ref{fig:CS_gen_samples_mean_photon_number_no_errors}). The mean photon number computed for each of the generated samples, setting $m=n=1$ in Eq.~(\ref{eq:mean_ph_num_Wunsche}) is denoted by a black circle in Fig.~\ref{fig:CS_gen_samples_mean_photon_number_no_errors}. This is within $4\%$ of the expected value for most output samples. As in the case of photon number states, the generation process produced some spurious samples. 

\begin{figure}
\centering
\includegraphics[width=0.45\textwidth]{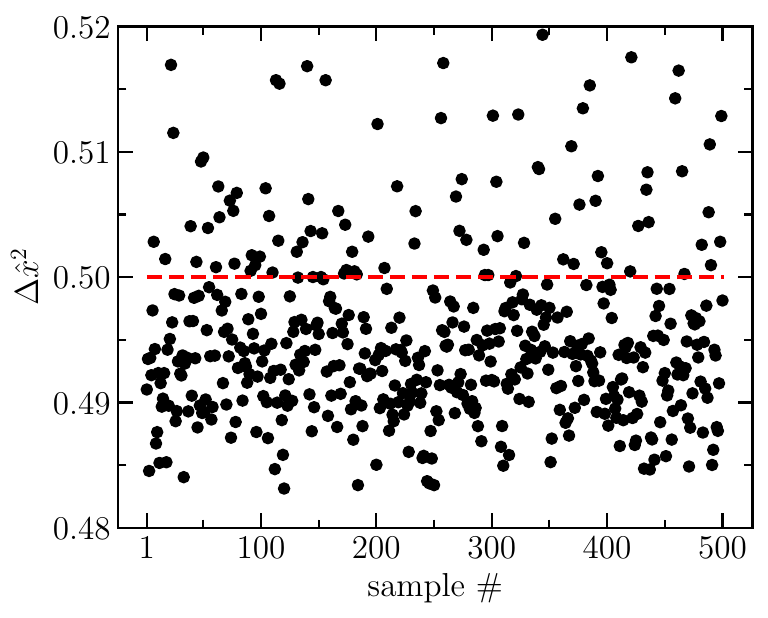}
\caption{The quadrature variance $\Delta \hat{x}^{2}$ (black circles) for the CS $|\alpha\rangle$ corresponding to each of the generated samples, along the $x$-quadrature. This is evaluated using the reconstructed PDFs along $\theta=0$. The horizontal red dashed line represents the theoretical value $0.5$, independent of the numerical value of $\alpha^{2}$. As in the case of Fock states, most of the black circles (generated values of the variance) lie within $4\%$ of $0.5$. The skewness about the red dashed horizontal line is discussed in Sec.~\ref{sec:nonlinear_cmap_CS}.}
\label{fig:variance_gen_image_CS_500_samples_theta_0.00000_no_errors}
\end{figure}
We now turn to the quadrature variances. The CS is a minimum uncertainty state equalizing the Heisenberg inequality, i.e., both the $x$ and the $p$ quadrature (equivalently, other canonically conjugate quadrature pairs) variances equal $0.5$, independent of the value of $\alpha$. In Fig.~\ref{fig:variance_gen_image_CS_500_samples_theta_0.00000_no_errors} we have presented a plot of the variance along the $x$-quadrature versus sample number for $500$ randomly selected output samples. Each sample could have one of the five possible values of $\alpha$ that we have used in the training program. As in the case of Fock states, most of the black circles (generated values of the variance) lie within $4\%$ of $0.5$. We notice from Figs.~\ref{fig:variance_gen_image_CS_500_samples_theta_0.00000_no_errors} and~\ref{fig:variance_gen_image_CS_500_samples_theta_1.57080_no_errors} (the analogous plot in the canonically conjugate $p$-quadrature) that there is a skewness in the distribution of black circles about the horizontal red dashed line. It is therefore necessary to investigate if this asymmetry can be minimized by a judicious choice of colormap, or whether it is an artifact of the generation process, whose genesis lies in the network architecture. We proceed to examine this aspect in the next section. 

\subsection{Choice of colormaps\label{sec:nonlinear_cmap_CS}}
\begin{figure}
\centering
\includegraphics[width=0.45\textwidth]{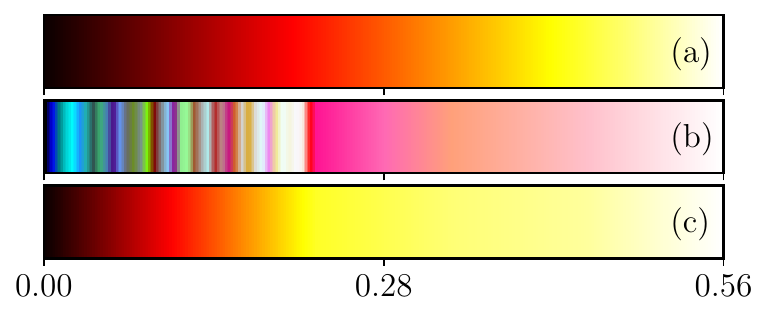}
\caption{Specification of the colormaps that were used in both the training and the reconstruction program. They are (a) the linear sequential, (b) the nonlinear and (c) the nonlinear sequential colormap. The colormap (b) has several disparate colors contiguously placed for values $\leqslant 0.28$ approximately, ordered in decreasing order of the RGB code (from pink ($\sim 0.28$) to black which is $0$). This particular choice is motivated by the need to capture the fall off in the tail of the reconstructed PDFs better than the default colormap (a). However, this necessitates the use of a regulator (see Eq.~(\ref{eq:regulator})) as (b) uses fewer colors ($\sim 45$) compared to $255$ colors used in (a) and (c). The colormap (c) is obtained by changing the spacing of colors in (a). Compared to (a) there are more shades in the range $0$ to $0.28$ in (c). We have verified that (b) and (c) are comparable in their performance, and therefore we have reported results corresponding only to the colormaps (a) and (b), in Sec.~\ref{sec:nonlinear_cmap_CS}.}
\label{fig:hot_nlin_nlin_hot_colormap_panel}
\end{figure}

As in the preceding section, we now consider input tomograms for the CS which are error-free. 
For illustrative purposes, we consider the CS $|\alpha\rangle$ with $\alpha = 0$ and $\sqrt{0.3}$ 
(Gaussians and displaced Gaussians). It is evident that the errors in the computed values of the mean photon number and the quadrature variances can be traced back to a mismatch between the input and the generated PDFs. This could possibly arise because the machine is unable to decipher intensity transitions in certain colormaps. To examine this aspect better, we have worked with three different types of colormaps using the same colormap for both the input and output tomograms. These are (a) the linear sequential, (b) the nonlinear and (c) the nonlinear sequential colormap. These colormaps are given in Fig.~\ref{fig:hot_nlin_nlin_hot_colormap_panel}. 

\begin{figure}
\centering
\includegraphics[width=0.45\textwidth]{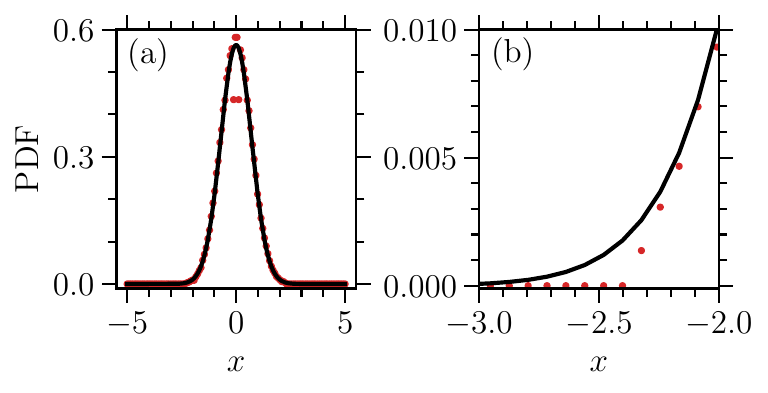}
\caption{Left to right: (a) Normalized reconstructed PDF (red dots) along the $x$-quadrature corresponding to a randomly selected output state $|0\rangle$. Reconstruction was carried out with a linear colormap. (b) The zoomed-in left tail of the PDF. The black line in both plots denotes the ideal Gaussian PDF along the $x$-quadrature, with mean zero and variance $0.5$. The red dots in (a) near $x = 0$ which are not on the Gaussian, do not significantly affect the moments of $\hat{x}$, since the small numerical value of $x$ diminishes their contribution. The sharper fall-off of the reconstructed PDF in comparison to the true PDF, particularly near the left tail in (b) and similarly near the right tail, is the primary reason for the departure of the variance from $0.5$.}
\label{gen_PDF_Fock_0_best_theta_0.00000_no_errors_panel}
\end{figure}

We first consider a normalized reconstructed PDF of the $0$-photon state (Fig.~\ref{gen_PDF_Fock_0_best_theta_0.00000_no_errors_panel}), generated for a specific output tomogram using a sequential linear colormap. The solid black lines are the ideal Gaussian PDF in the $x$-quadrature (Fig.~\ref{gen_PDF_Fock_0_best_theta_0.00000_no_errors_panel}(a)) and its zoomed-in version (Fig.~\ref{gen_PDF_Fock_0_best_theta_0.00000_no_errors_panel}(b)). The red dots define the reconstructed PDF (Fig.~\ref{gen_PDF_Fock_0_best_theta_0.00000_no_errors_panel}(a)) and its zoomed-in version near the left tail of the Gaussian (Fig.~\ref{gen_PDF_Fock_0_best_theta_0.00000_no_errors_panel}(b)). The red dots in Fig.~\ref{gen_PDF_Fock_0_best_theta_0.00000_no_errors_panel}(a) close to $x=0$ which do not lie on the Gaussian, do not contribute significantly to the moments, as these points are weighted by small values of $x$. 

However, the errors during the reconstruction process are significant in the neighborhood of points where the slope of the PDF changes rapidly, for instance close to the extrema, and the transition to the tail. It is evident that the contributions arising due to the points in this region, which do not coincide with the ideal Gaussian tail, could get amplified when moments are calculated. A similar situation prevails near the right tail of the Gaussian. This problem arises in other quadratures as well. 

We note that $|0\rangle$ was merely one of the states among a set of Fock states $|n\rangle$ ($n=0,1,2,\dots,5$) used for training, and hence the corresponding PDFs were a collection of unimodal and multimodal distributions. In order to examine if the errors listed above are reduced if the training is carried out exclusively with Gaussians, we have also considered the output PDFs in the case of coherent states. We recall that the input in this case had tomograms corresponding to $\alpha = 0, \sqrt{0.1}, \sqrt{0.3}, \sqrt{0.5}$ and $1$. In particular, we have investigated if errors arise in the reconstructed PDFs for $\alpha=0$, in order that comparison can be readily made with the results in the case of the $0$-photon state. 

\begin{figure}
\centering
\includegraphics[width=0.45\textwidth]{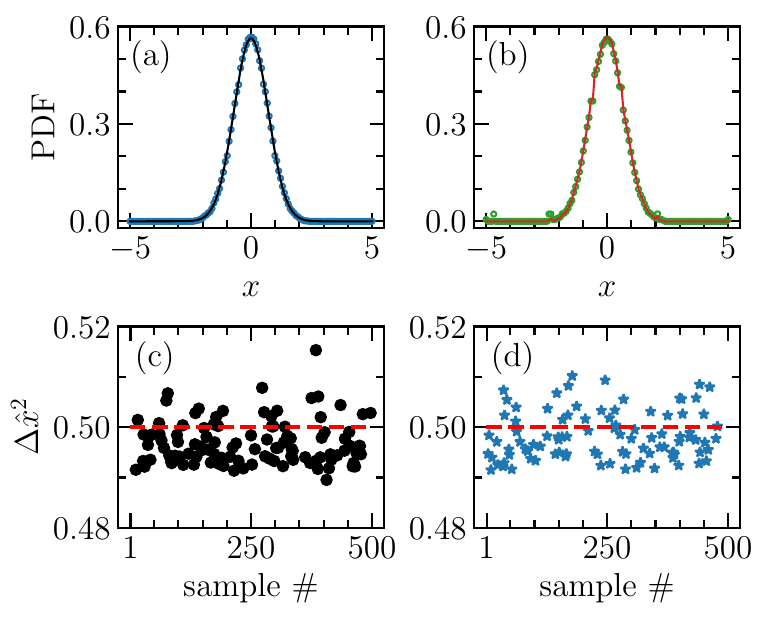}
\caption{Top panel: The normalized reconstructed PDF in the $x$-quadrature for the CS ($\alpha = 0$), with (a) a sequential linear colormap (black line: true PDF, and open blue circles: reconstructed PDF). (b) A nonlinear colormap without the regulator (green open circles) and with the regulator setting $x_{0}=0$, $L=2.3$ and $m=5$ in Eq.~(\ref{eq:regulator}) (red line). \\
Bottom panel: The corresponding quadrature variances $\Delta \hat{x}^{2}$ for all generated samples with the red dashed horizontal line indicating the vacuum variance set at $\Delta \hat{x}^{2} = 0.5$. (c) The sequential linear colormap (a black circle for each sample). $106$ circles with $18$ above $0.5$. (d) The nonlinear colormap with the regulator (a blue asterisk for each sample). $96$ asterisks with $27$ above $0.5$. The asymmetry about the red horizontal dashed line in (c) and (d) is inherent to the generation process. }
\label{fig:gen_PDF_variance_theta_0.00000_alpha_0.00000_seqlin_nlin_cmap_no_errors_panel}
\end{figure}

In Fig.~\ref{fig:gen_PDF_variance_theta_0.00000_alpha_0.00000_seqlin_nlin_cmap_no_errors_panel}(a) the normalized reconstructed PDF  in the $x$-quadrature corresponding to a randomly selected output tomogram with $\alpha=0$ is shown. The black solid line is the ideal PDF for $\alpha=0$ and the blue open circles correspond to the reconstructed PDF. A zoomed-in version of Fig.~\ref{fig:gen_PDF_variance_theta_0.00000_alpha_0.00000_seqlin_nlin_cmap_no_errors_panel}(a) (not shown here) is similar to Fig.~\ref{gen_PDF_Fock_0_best_theta_0.00000_no_errors_panel}(b). A plot of the variance in the $x$-quadrature computed from the reconstructed PDF versus sample number is shown in Fig.~\ref{fig:gen_PDF_variance_theta_0.00000_alpha_0.00000_seqlin_nlin_cmap_no_errors_panel}(c). The skewness observed in the distribution of the black circles for all values of $\alpha$, about the horizontal red line in Figs.~\ref{fig:variance_gen_image_CS_500_samples_theta_0.00000_no_errors} and~\ref{fig:variance_gen_image_CS_500_samples_theta_1.57080_no_errors}, is also reflected in this plot (for $\alpha = 0$ alone). Out of a total of $106$ black circles, only $18$ lie above the horizontal red dashed line. 

In order that the machine learns the tail of the input PDFs accurately,we now investigate the consequences of using a nonlinear colormap instead of the sequential linear colormap. In the nonlinear colormap, the color definitions in the tail of the PDFs in various quadratures are enhanced by increasing the number of shades of the different colors used, nonlinearly. This is in contrast to the sequential linear colormap where the range of values of the tomogram $w(X_{\theta}, \theta)$ for a given state (with $(X_{\theta}, \theta)$ defining a pixel) is associated with a range of colors, both ranges being equispaced. The nonlinear colormap facilitates tracking the transition from a finite small value of the tomogram smoothly to zero, more efficiently. 

\begin{figure}
\centering
\includegraphics[width=0.45\textwidth]{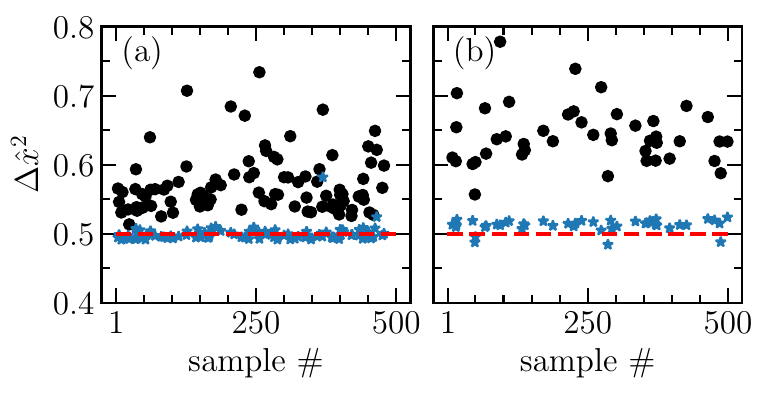}
\caption{The quadrature variance $\Delta \hat{x}^{2}$ corresponding to (a) $\alpha=0$ and (b) $\alpha=\sqrt{0.3}$. The red dashed line is the theoretical (expected) variance $0.5$. Each black circle (blue asterisk) is the variance for one randomly generated sample without the regulator (with the regulator). It is clear that the regulator helps reduce the contribution of spurious values (primarily originating at the tail of the generated PDFs), and brings the variance close to $0.5$.}
\label{fig:variance_CS_alpha_0.00000_0.54772_theta_0.00000_nlin_cmap_panel_no_errors_panel}
\end{figure}

A plot of a generated sample PDF (green open circles) in the $x$-quadrature with $\alpha = 0$, reconstructed using the nonlinear colormap is shown in Fig.~\ref{fig:gen_PDF_variance_theta_0.00000_alpha_0.00000_seqlin_nlin_cmap_no_errors_panel}(b). However, we note that there are green open circles lying outside the tail of the Gaussian. This can lead to errors in the observables calculated from the PDF. Therefore, we have weighted the PDF with a regulator function
\begin{equation}
f(x) = \exp [-((x-x_{0})/L)^{2s}],
\label{eq:regulator}
\end{equation}
where $x_{0}$ and  $L$ are the shift and the cut-off values respectively, and $s$ is the parameter that controls the fall-off. The role of the regulator is to exponentially suppress values exceeding $L$ (centered about $x_{0}$) and as a result it effectively removes the spurious points in the tail. A larger value of $s$ signifies faster exponential fall-off and vice versa. Therefore, in what follows, we first used Eq.~(\ref{eq:regulator}) to suppress the spurious values of the generated PDFs. The generic values of $L$ and $s$ were found to be $2.3$ and $5$, respectively. In Fig.~\ref{fig:gen_PDF_variance_theta_0.00000_alpha_0.00000_seqlin_nlin_cmap_no_errors_panel}(b) the regulated PDF is given by the red solid line. All the regulated PDFs were normalized and the moments were computed from them. 

The variances obtained in the $x$-quadrature for $\alpha = 0$ are shown by black circles for the sequential colormap (Fig.~\ref{fig:gen_PDF_variance_theta_0.00000_alpha_0.00000_seqlin_nlin_cmap_no_errors_panel}(c)), and by blue asterisks for the nonlinear colormap with the regulator (Fig.~\ref{fig:gen_PDF_variance_theta_0.00000_alpha_0.00000_seqlin_nlin_cmap_no_errors_panel}(d)). The red dashed line denotes the expected variance for this state. While the skewness in Fig.~\ref{fig:gen_PDF_variance_theta_0.00000_alpha_0.00000_seqlin_nlin_cmap_no_errors_panel}(d) is marginally reduced compared to that in Fig.~\ref{fig:gen_PDF_variance_theta_0.00000_alpha_0.00000_seqlin_nlin_cmap_no_errors_panel}(c), the performances of both colormaps are similar, provided the regulator is used in the case of the nonlinear colormap. 
 
The role played by the regulator is seen clearly in Figs.~\ref{fig:variance_CS_alpha_0.00000_0.54772_theta_0.00000_nlin_cmap_panel_no_errors_panel}(a) and~(b) corresponding to $\alpha = 0$ and $\sqrt{0.3}$ respectively. The black circles in both figures are the variances in the $x$-quadrature computed with the nonlinear colormap, without the regulator, and the blue asterisks correspond to computation with the regulator. It is evident that the regulator has reduced the contribution from the spurious points near the tail of the Gaussian and that the variance is now given by a small spread about $0.5$. Therefore, the nonlinear colormap fares reasonably well only if the regulator is used. 

To quantify the dissimilarities between the true PDF and the reconstructed PDF using (1) the sequential linear colormap, (2) the nonlinear colormap without the regulator, and (3) the nonlinear colormap with the regulator, the Wasserstein distance $W_{1}$ (defined in footnote~\ref{fn:wass_dist_bw_pdf}) between these two PDFs has been computed in these cases. The variance along the $x$-quadrature for the sample PDF selected for this purpose is $0.49$, $0.57$ and $0.49$, for cases (1), (2) and (3), respectively. The corresponding values of $W_{1}$ are $0.01$, $0.03$ and $0.01$. This corroborates the inference that the nonlinear colormap without the regulator does not fare well and that the sequential linear colormap and the nonlinear colormap with the regulator, are comparable in their performance. We have also verified that replacing the sequential map discussed above with an appropriate sequential nonlinear map does not change the results significantly. 

Finally, we remark that the skewness about the expected variance both in the case of the $0$-photon state and the CS seems to be inherent in the generation process, due to the specific architecture that has been used. However, the computed numerical values of the physical observables such as the mean photon number and the variance are within $4\%$ of their true values in most cases. These computations have been carried out in a straightforward manner from the tomograms themselves. 

The foregoing analysis has also been carried out with $256 \times 256$ pixels in the tomograms. The inferences drawn from this calculation are essentially similar to those reported earlier. We have verified that the skewness is present even if the computations are carried out with raw tomographic data, i.e., numerical arrays (Appendix~\ref{sec:appendix_numerical_arrays}) instead of images (RGB data). The genesis of the skewness therefore seems to lie in the architecture that has been used. Hence, in the following sections also, the investigations reported correspond to $128 \times 128$ pixels. We now proceed to examine the efficacy of the training program when a photon is added to the CS. 

\subsection{Single-photon added CS\label{sec:PACS}}
\begin{figure}
\centering
\includegraphics[width=0.45\textwidth]{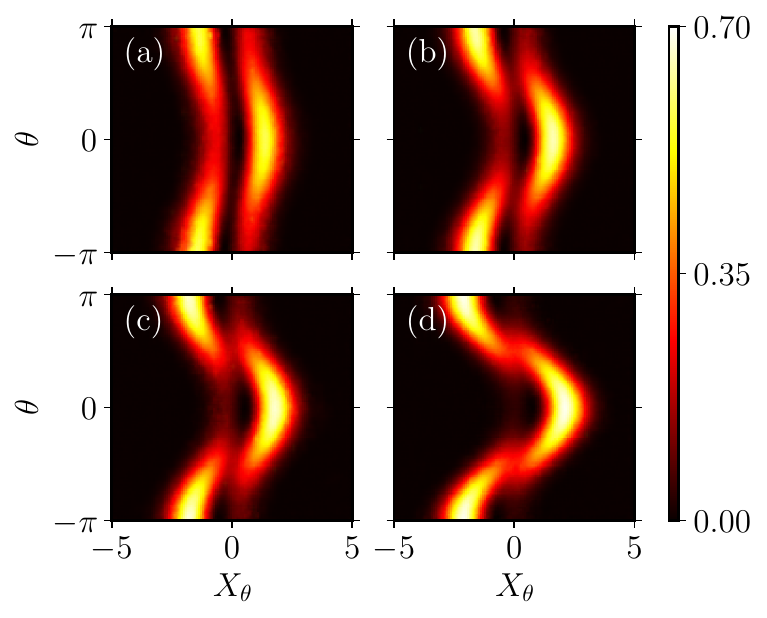}
\caption{Representative generated samples for the $1$-PACS $|\alpha,1\rangle$ obtained after training with the WGAN algorithm for $25000$ epochs. Here $\alpha=$ $\sqrt{0.1}$, $\sqrt{0.3}$, $\sqrt{0.5}$ and $1$ in (a), (b), (c) and (d), respectively. It can be seen that the vertical cut becomes blurred with increase in the value of $\alpha$. This is consistent with Fig.~\ref{fig:tomogram_PACS_m_1_g1_alpha_X_theta_-5_5_panel} (ideal tomograms and caption).}
\label{fig:generated_PACS_tomograms_no_errors_panel}
\end{figure}
\begin{figure}
\centering
\includegraphics[width=0.45\textwidth]{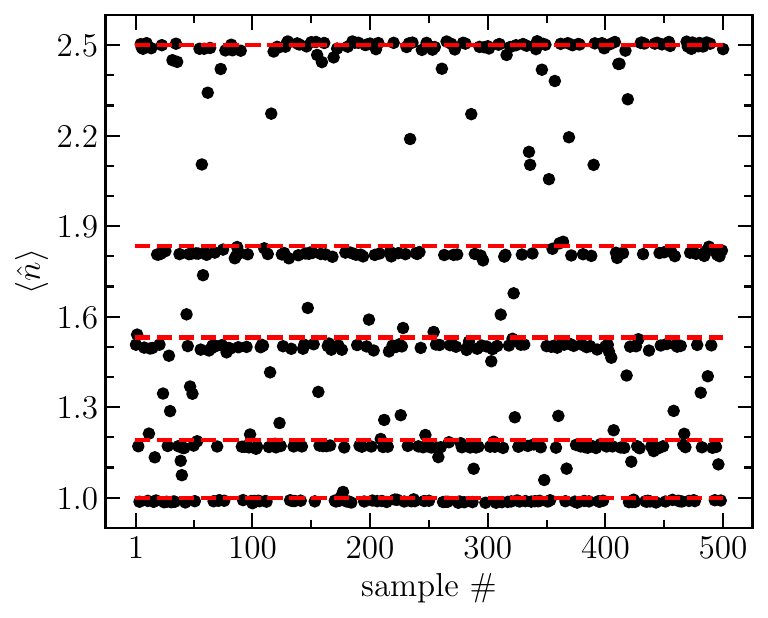}
\caption{The mean photon number $\langle \hat{n} \rangle$ (black circles) for the $1$-photon added CS $|\alpha,1\rangle$ corresponding to each of the generated samples. This is evaluated using the reconstructed PDFs along $\theta=0, \pi/3$ and $2\pi/3$, and the expression in Eq.~(\ref{eq:mean_ph_num_Wunsche}) setting $m=n=1$. The horizontal red dashed lines represent the theoretical values of $\langle \hat{n} \rangle$ given by $1, 1.191, 1.531, 1.833$, and $2.5$, corresponding to $|\alpha|^{2}=0, 0.1, 0.3, 0.5$ and $1$, respectively.}
\label{fig:PACS_gen_samples_mean_photon_number_no_errors}
\end{figure}
At the end of the training program $500$ generated samples were randomly selected for characterization. Representative samples are shown in Fig.~\ref{fig:generated_PACS_tomograms_no_errors_panel}. The expectation is that a considerable number of output tomograms should correspond to $|\alpha, 1\rangle$ with $\alpha = 0.0, \sqrt{0.1}, \sqrt{0.3}, \sqrt{0.5}$ and $1.0$, as these were the values used for training.

As in earlier cases the mean photon number for all the $500$ selected samples was computed using Eq.~(\ref{eq:mean_ph_num_Wunsche}) (black circles in Fig.~\ref{fig:PACS_gen_samples_mean_photon_number_no_errors}). As stated earlier this is given by $2L_{2}(-|\alpha|^{2})/L_{1}(-|\alpha|^{2}) - 1$, where $L_{m}$ is the Laguerre polynomial of order $m$. These values are denoted by red horizontal dashed lines in Fig.~\ref{fig:PACS_gen_samples_mean_photon_number_no_errors} for different values of $\alpha$. We see that with increase in $\langle \hat{n} \rangle$ there in an increase in the number of outliers. This is a consequence of the fact that the number of maxima and minima, and hence the number of color gradients in the PDFs increase with $\alpha$. 

\begin{figure}
\centering
\includegraphics[width=0.45\textwidth]{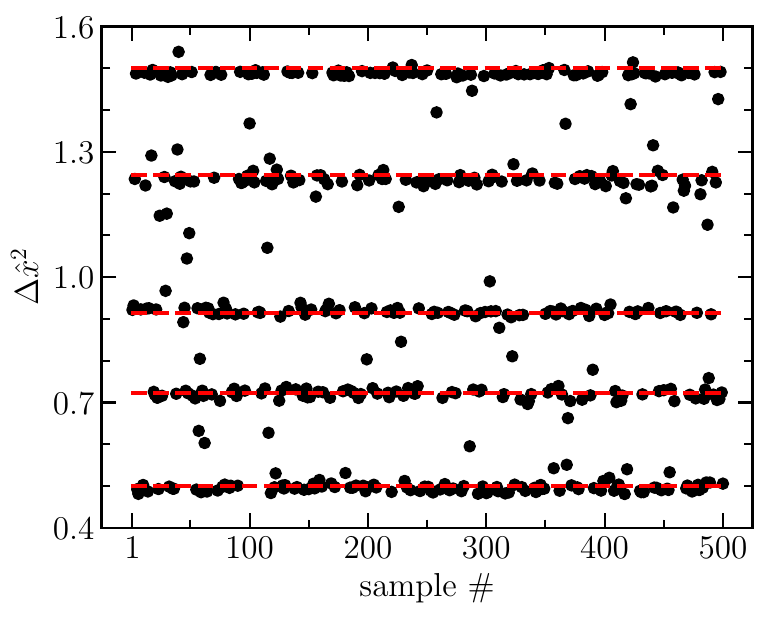}
\caption{The quadrature variance $\Delta \hat{x}^{2}$ (black circles) for the $1$-photon added CS $|\alpha,1\rangle$ corresponding to each of the generated samples, along the $x$-quadrature. This is evaluated using the reconstructed PDFs along $\theta=0$. The red horizontal dashed lines represent the theoretically computed values $1.50, 1.24, 0.92, 0.72$, and $0.50$, corresponding to $\alpha^{2}=0, 0.1, 0.3, 0.5$ and $1$, respectively.}
\label{fig:variance_gen_image_PACS_500_samples_theta_0.00000_no_errors}
\end{figure}
A similar exercise has been carried out for the quadrature variances $\Delta \hat{X}^{2}_{\theta}$ for different values of $\theta$. The variances corresponding to the $x$ and $p$-quadratures are shown in Figs.~\ref{fig:variance_gen_image_PACS_500_samples_theta_0.00000_no_errors} and~\ref{fig:variance_gen_image_PACS_500_samples_theta_1.57080_no_errors} respectively. It is clear that for most samples the computed variances are within reasonable error tolerance (within $4\%$) of the expected theoretical values. A small percentage of outliers can also be observed. These increase with the mean photon number as expected. 

In our studies we have allowed for $4\%$ error tolerance in the values of the mean photon number and quadrature variance computed from the generated tomograms of the various states. We have elaborated upon this choice of error tolerance in Appendix~\ref{sec:appendix_choice_of_err_tol}. 

Having generated tomograms corresponding to the CS and the single-photon added CS, and characterized them, we now proceed to examine if recent experimental results on classical and quantum amplification, involving these states, are reproduced successfully in the generation program.

\subsection{Comparison with experiment\label{sec:Fadrny_PACS_expt}}
\begin{figure}
\centering
\includegraphics[width=0.45\textwidth]{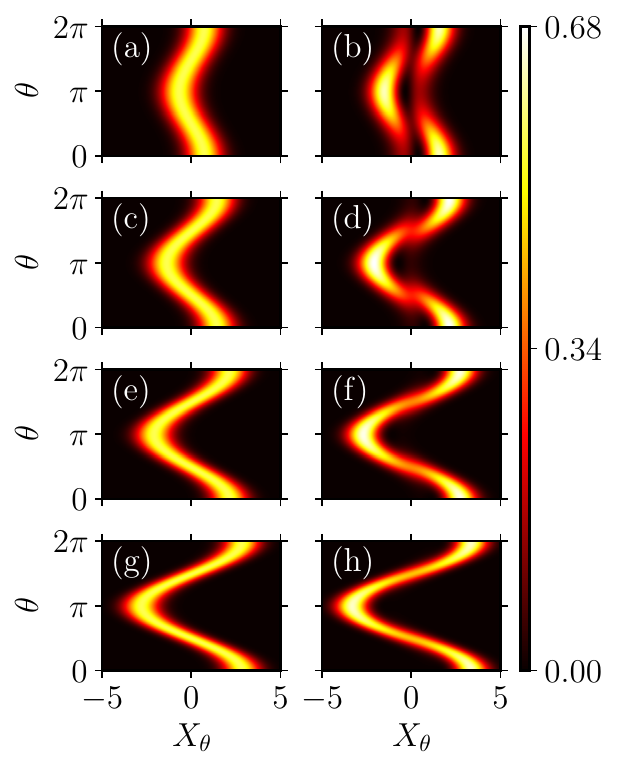}
\caption{Single-mode optical tomograms. Left column panels correspond to the amplified CS $|{\sf g}_{1}(\alpha)\alpha\rangle$, and right column panels to $|\alpha,1\rangle$. Here $\alpha=0.5$ for (a) and (b), $1.0$ for (c) and (d), $1.5$ for (e) and (f), and $2.0$ for (g) and (h) respectively. Note that for $\alpha > 1$, the vertical cut in the tomograms corresponding to $|\alpha,1\rangle$ is absent, making them visually appear like the tomogram for the CS [see (e) -- (h)]. Hence for $\alpha > 1$ the fidelity between $|{\sf g}_{1}(\alpha)\alpha\rangle$ and $|\alpha,1\rangle$ is maximized (experiment reported in~\cite{Fadrny:2024}). However, the variances along the $x$-quadrature for instance, computed directly from these tomograms are different for the CS $|{\sf g}_{1}(\alpha)\alpha\rangle$ and $|\alpha,1\rangle$ (see Fig.~\ref{fig:variance_PACS_m1_g1_alpha} and also~\cite{Soumyabrata:2025c}). \\
The tomograms represented here appear to be inverted about the vertical axis, compared to those in Figs.~\ref{fig:fig_tomogram_Fock_0_1_2_PACS_alpha_0.5_m_0_1_2_panel}(d) and~(e) for instance. This can be readily explained from the symmetry property in Eq.~(\ref{eq:w_singlemode_symmetry}) and the fact that the range along the $\theta$ axis is different in both figures.}
\label{fig:tomogram_PACS_m_1_g1_alpha_X_theta_-5_5_panel}
\end{figure}
In a recent experimental paper~\cite{Fadrny:2024}, the authors have considered conditional addition of one and two photons to a CS $|\alpha\rangle$, and shown that for a range of real values of $\alpha$, the resulting state is close in fidelity to an amplified  CS $|{\sf g}_{m}(\alpha)\alpha\rangle$, for $m=1, 2$. The amplification gain ${\sf g}_{m}(\alpha)$ is given by $\langle \alpha,m|\hat{a}|\alpha,m \rangle/\alpha$. This is borne out in the reported fidelity plots between $|{\sf g}_{m}(\alpha)\alpha\rangle$ and $|\alpha,m\rangle$, as functions of $\alpha$ (Fig.~4 of~\cite{Fadrny:2024}). Further, in Fig.~5 of~\cite{Fadrny:2024}, the authors have plotted the fidelity between a CS $|\beta_{\rm opt}\rangle$ and $|\alpha,m\rangle$, as a function of $\alpha$ with $\beta_{\rm opt} = \alpha[1+(1+4m/{\alpha}^{2})^{1/2}]/2$ ($\alpha$ real). (It can be established that $|\beta_{\rm opt}\rangle$ is the CS with maximum fidelity with $|\alpha,m\rangle$ for a fixed $\alpha$ and $m$. It is worth noting that for $\alpha \geqslant 1$, ${\sf g}_{m}(\alpha)\alpha \approx \beta_{\rm opt}$.) 

The fidelities were obtained in the experiment after reconstructing the appropriate coherent states and $|\alpha,1\rangle$ (equivalently, the corresponding density matrices and Wigner functions) from the experimental data. The findings of relevance to us are that the fidelity between $|{\sf g}_{1}(\alpha)\alpha\rangle$ and $|\alpha,1\rangle$ is close to unity for $\alpha \geqslant 1$. 

\begin{figure}
\centering
\includegraphics[width=0.45\textwidth]{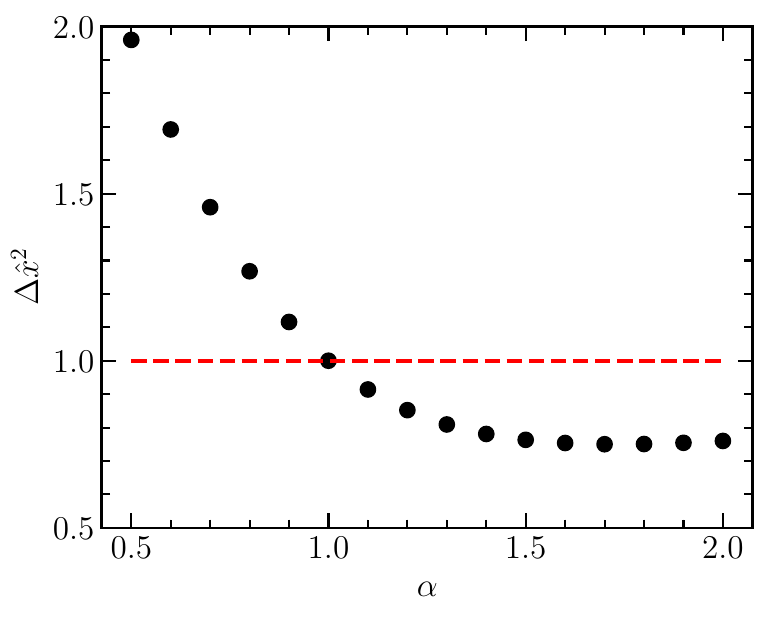}
\caption{The quadrature variance $\Delta {\hat x}^{2}$ (black circles) corresponding to the state $|\alpha,1\rangle$, computed directly from the tomograms for $0.5 \leqslant \alpha \leqslant 2.0$. The horizontal red dashed line is the quadrature variance for the CS $|{\sf g}_{1}(\alpha)\alpha\rangle$ computed directly from the tomograms corresponding to $0.5 \leqslant \alpha \leqslant 2.0$. As expected this is a constant. It has been set to unity for ready comparison with the experimental report in~\cite{Fadrny:2024}. For $\alpha = 1$, $\Delta {\hat x}^{2}$ for both $|{\sf g}_{1}(\alpha)\alpha\rangle$ and $|\alpha,1\rangle$ are equal. For $\alpha > 1$, $\Delta \hat{x}^{2} < 1$ for $|\alpha,1\rangle$ indicating that $|\alpha,1\rangle$ is squeezed along the $x$-quadrature. Thus, the amplified CS does not display squeezing properties but the state resulting due to photon addition to the CS shows quantum amplification.}
\label{fig:variance_PACS_m1_g1_alpha}
\end{figure}

From the tomographic point of view this implies that the tomograms corresponding to these two states would appear very similar visually. In Fig.~\ref{fig:tomogram_PACS_m_1_g1_alpha_X_theta_-5_5_panel} we have demonstrated that the vertical cut that appears in the tomogram for $|\alpha,1\rangle$ ($\alpha < 1$) disappears for $\alpha > 1$. It is therefore important to compute quantum numbers corresponding to the amplified CS and the $1$-PACS to distinguish between them. We have carried this out by computing variances with ideal tomograms of both states for the purpose of demonstration (Fig.~\ref{fig:variance_PACS_m1_g1_alpha}). It is clear from the plot of $\Delta \hat{x}^{2}$ versus $\alpha$ that the variance corresponding to $|\alpha,1\rangle$ (black circles) is squeezed for $\alpha > 1$, in contrast to the variance for the CS $|{\sf g}_{1}(\alpha)\alpha\rangle$ which is a constant independent of $\alpha$ (scaled to $1$: red dashed horizontal line). This distinguishes between the two states. 

\begin{figure}[!ht]
\centering
\includegraphics[width=0.45\textwidth]{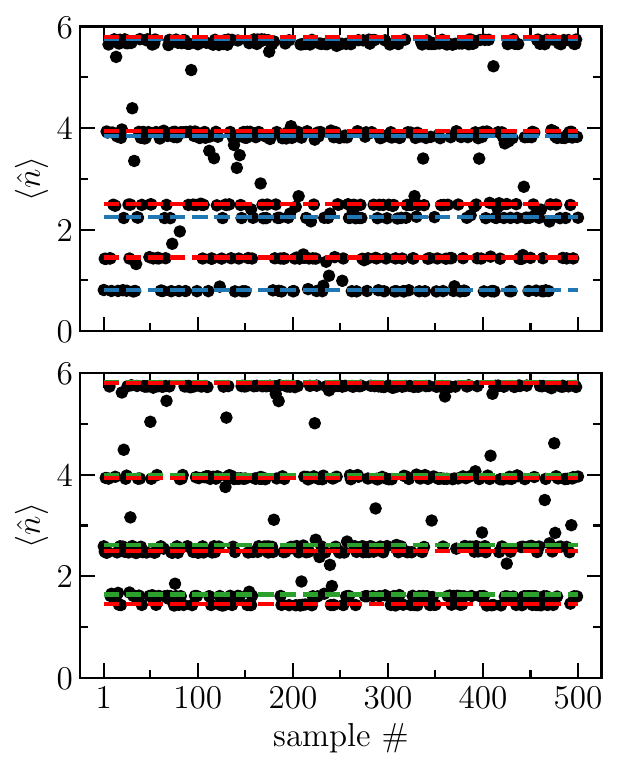}
\caption{The theoretical values of $\langle \hat{n} \rangle = ({\sf g}_{1}(\alpha)\alpha)^{2}$ for the amplified CS $|{\sf g}_{1}(\alpha)\alpha\rangle$ are $0.81, 2.25, 3.84$ and $5.76$ corresponding to $\alpha = 0.5, 1.0, 1.5$ and $2.0$ respectively (blue horizontal dashed lines in the top panel). The theoretical values of $\langle \hat{n} \rangle = \beta_{\rm opt}^{2}$ for the optimal CS $|\beta_{\rm opt}\rangle$ are $1.64, 2.62, 4$ and $5.83$ corresponding to the same values of $\alpha$ as above (green horizontal dashed lines in the bottom panel). \\ 
In both panels the red horizontal dashed lines denote the theoretical values of $\langle \hat{n} \rangle$ given by $1.45, 2.5, 3.94$ and $5.8$, corresponding to $|\alpha,1\rangle$, for the same values of $\alpha$ as above. The input (and hence the output) samples comprise tomograms corresponding to both $|{\sf g}_{1}(\alpha)\alpha\rangle$ and $|\alpha,1\rangle$ in the top panel, and $|\beta_{\rm opt}\rangle$ and $|\alpha,1\rangle$ in the bottom panel. Each black circle represents the value of $\langle \hat{n} \rangle$ for one of the $500$ randomly generated output tomograms in both panels. This is evaluated using the reconstructed PDFs along $\theta=0, \pi/3$ and $2\pi/3$, and the expression in Eq.~(\ref{eq:mean_ph_num_Wunsche}) setting $m=n=1$. Most of the black circles are in the neighborhood of either the red or the blue lines in the top panel, and red or the green lines in the bottom panel. This indicates that values of $\langle \hat{n} \rangle$ corresponding to the generated samples of $|{\sf g}_{1}(\alpha)\alpha\rangle$, $|\beta_{\rm opt}\rangle$ and $|\alpha,1\rangle$ agree well with their theoretical values. We see that with increase in $\alpha$ the black circles as well as the corresponding red and blue lines (top panel), and red and green lines (bottom panel) move close to each other, and for $\alpha > 1.0$, lie almost on top of each other. This indicates that for large values of $\alpha$, an amplified CS (respectively, an optimal CS) and a photon added CS have similar characteristics as reported in the experiment~\cite{Fadrny:2024}.} 
\label{fig:PACS_gen_samples_mean_photon_number_g1alpha_alpha_beta_opt_no_errors_panel}
\end{figure}

\begin{figure}[!ht]
\centering
\includegraphics[width=0.45\textwidth]{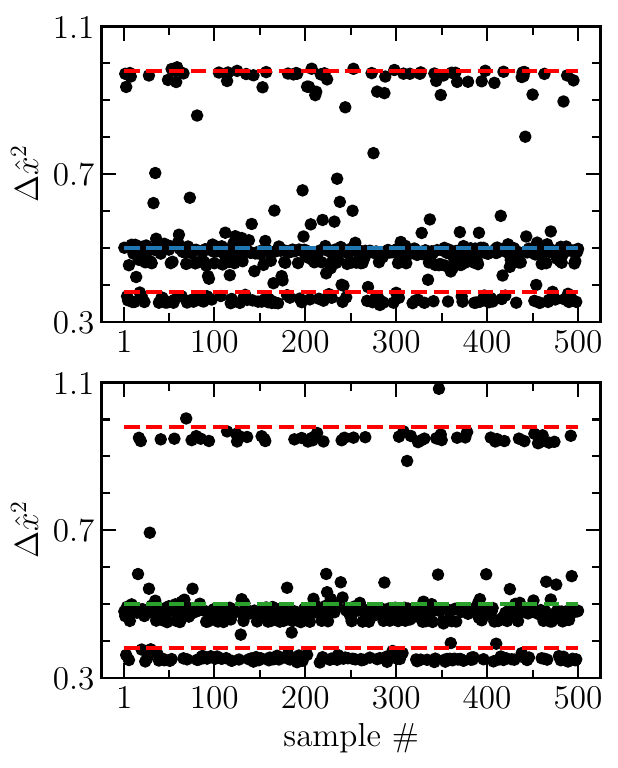}
\caption{The theoretical value ($0.5$) of the quadrature variance $\Delta \hat{x}^{2}$ is shown (blue horizontal dashed line) in the top panel for the amplified CS $|{\sf g}_{1}(\alpha)\alpha\rangle$, and in the bottom panel (green horizontal dashed line) for the optimal CS $|\beta_{\rm opt}\rangle$. The values of $\alpha$ used for the training program are $0.5, 1.0, 1.5$ and $2.0$. \\  
In both panels the red horizontal dashed lines denote the theoretical values of $\Delta \hat{x}^{2}$ given by $0.98$ ($\alpha=0.5$), $ 0.5$ ($\alpha = 1.0$), $0.382$ ($\alpha=1.5$) and $0.38$ ($\alpha=2.0$), corresponding to $|\alpha,1\rangle$. $\Delta \hat{x}^{2} = 0.38$ and $0.382$ indicates squeezing. The corresponding red horizontal dashed lines are not distinguishable in the chosen scale. The input (and hence the output) samples comprise tomograms for both $|{\sf g}_{1}(\alpha)\alpha\rangle$ and $|\alpha,1\rangle$ in the top panel, and $|\beta_{\rm opt}\rangle$ and $|\alpha,1\rangle$ in the bottom panel. Each black circle represents the value of $\Delta \hat{x}^{2}$ for one of the $500$ randomly generated output tomograms in both panels. This is evaluated using the reconstructed PDFs along $\theta=0$. Most of the black circles are in the neighborhood of either the red or the blue lines in the top panel, and red or the green lines in the bottom panel. This indicates that values of $\Delta \hat{x}^{2}$ corresponding to the generated samples of $|{\sf g}_{1}(\alpha)\alpha\rangle$, $|\beta_{\rm opt}\rangle$ and $|\alpha,1\rangle$ agree well with their theoretical values.}
\label{fig:variance_gen_image_PACS_500_samples_theta_0.00000_g1alpha_alpha_beta_opt_no_errors_panel}
\end{figure}

To ascertain the extent of success of our ML procedure, we have examined whether the mean photon numbers corresponding to $|{\sf g}_{1}(\alpha)\alpha\rangle$, $|\alpha,1\rangle$ and $|\beta_{\rm opt}\rangle$ are close to each other for $\alpha \geqslant 1$. Two separate training programs were carried out with tomograms corresponding to (a) $|\alpha,1\rangle$ and $|{\sf g}_{1}(\alpha)\alpha\rangle$, and (b) $|\alpha,1\rangle$ and $|\beta_{\rm opt}\rangle$. The learning process was for values of $\alpha$ which the experimenters had considered, namely, $0.5, 1.0, 1.5$ and $2.0$. In both cases (a) and (b), the machine was trained using the WGAN algorithm for $25000$ epochs, and the PDFs corresponding to the generated output tomograms were reconstructed. From the PDFs, $\langle \hat{n} \rangle$ and $\Delta \hat{X}_{\theta}^{2}$ along specific quadratures were computed. The mean photon number and the quadrature variance corresponding to the states considered in (a) and (b) are shown in Figs.~\ref{fig:PACS_gen_samples_mean_photon_number_g1alpha_alpha_beta_opt_no_errors_panel} and~\ref{fig:variance_gen_image_PACS_500_samples_theta_0.00000_g1alpha_alpha_beta_opt_no_errors_panel} respectively, for $\alpha = 0.5, 1.0, 1.5$ and $2.0$. In the top panel of Fig.~\ref{fig:PACS_gen_samples_mean_photon_number_g1alpha_alpha_beta_opt_no_errors_panel} the theoretical values of $\langle \hat{n} \rangle$ for $|{\sf g}_{1}(\alpha)\alpha\rangle$ given by $({\sf g}_{1}(\alpha)\alpha)^{2}$ (indicated by blue horizontal dashed lines) are equal to $0.81$ ($\alpha=0.5$), $2.25$ ($\alpha=1.0$), $3.84$ ($\alpha=1.5$) and $5.76$ ($\alpha=2.0$). In the bottom panel the corresponding values for $|\beta_{\rm opt}\rangle$ are given by $\beta_{\rm opt}^{2}$ (green horizontal dashed lines). These are $1.64$ ($\alpha=0.5$), $2.62$ ($\alpha=1.0$), $4$ ($\alpha=1.5$) and $5.83$ ($\alpha=2.0$). In both panels the red horizontal dashed lines denote the theoretical values of $\langle \hat{n} \rangle$ corresponding to $|\alpha,1\rangle$, for the same values of $\alpha$ as above. These are $1.45, 2.5, 3.94$ and $5.8$. 

The input (and hence the output) samples comprise tomograms corresponding to both $|{\sf g}_{1}(\alpha)\alpha\rangle$ and $|\alpha,1\rangle$ (top panel), and $|\beta_{\rm opt}\rangle$ and $|\alpha,1\rangle$ (bottom panel). From the generated tomograms $500$ samples were randomly selected (for each panel), and the corresponding values of $\langle \hat{n} \rangle$ (represented by black circles) were computed from the reconstructed PDFs along $\theta=0, \pi/3$ and $2\pi/3$, using Eq.~(\ref{eq:mean_ph_num_Wunsche}). Most of these circles are in the vicinity of either the red or the blue lines in the top panel, and red or the green lines in the bottom panel, indicating that the training program has been successful. 

We now proceed to discuss the findings pertaining to the quadrature variance. The $500$ generated samples that were selected to obtain the mean photon number, were also used to calculate $\Delta \hat{x}^{2}$. The theoretical value ($0.5$) for the variance of both the amplified CS $|{\sf g}_{1}(\alpha)\alpha\rangle$ and the optimal CS $|\beta_{\rm opt}\rangle$ are denoted by the blue (green) horizontal dashed line in the top (bottom) panel of Fig.~\ref{fig:variance_gen_image_PACS_500_samples_theta_0.00000_g1alpha_alpha_beta_opt_no_errors_panel}. In the case of the $1$-PACS, $\Delta \hat{x}^{2}$ decreases with increase in $\alpha$. In both panels the corresponding values are denoted by the red horizontal dashed lines with $\Delta \hat{x}^{2} = 0.98$ ($\alpha=0.5$), $0.5$ ($\alpha = 1.0$), $0.382$ ($\alpha=1.5$) and $0.38$ ($\alpha=2.0$). This state is squeezed, i.e., $\Delta \hat{x}^{2} < 0.5$ for $\alpha > 1.0$. The red lines corresponding to $\Delta \hat{x}^{2} = 0.382$ and $0.38$ are not distinguishable in the chosen scale. 

Each value of $\Delta \hat{x}^{2}$ obtained from the reconstructed PDF along $\theta=0$ is represented by a black circle in both panels of Fig.~\ref{fig:variance_gen_image_PACS_500_samples_theta_0.00000_g1alpha_alpha_beta_opt_no_errors_panel}. As in Fig.~\ref{fig:PACS_gen_samples_mean_photon_number_g1alpha_alpha_beta_opt_no_errors_panel}, most of the black circles lie in the neighborhood of the theoretical values corresponding to the different states. We therefore infer that characterization of these states in terms of the mean photon number and the quadrature variance has been reasonably successful. Further, we see that with increase in $\alpha$ the generated values of the mean photon numbers corresponding to $|{\sf g}_{1}(\alpha)\alpha\rangle$ and $|\beta_{\rm opt}\rangle$ move close to the mean photon number for $|\alpha,1\rangle$. This indicates that for sufficiently large $\alpha$, both amplified and optimal coherent states have characteristics which are similar to those of the $1$-PACS. This is consistent with the results reported in the literature~\cite{Fadrny:2024}. We discuss further about classification of these states in the next section. 

\subsection{Classification\label{sec:classification}}
An important aspect of machine learning for tomograms is that the mean photon number, the variance in various quadratures, and the higher-order moments can be readily computed from the PDFs that comprise the tomograms. This suffices to distinguish between different classes of states. 

For instance, consider the case when the training set consists solely of the photon number states $|n\rangle$ for different values of $n$ (Sec.~\ref{sec:Fock}). If, a priori it is known that the set of generated samples is expected to belong to this class of states, computation of the mean photon number (Fig.~\ref{fig:Fock_gen_samples_mean_photon_number_no_errors}) or the quadrature variances (Figs.~\ref{fig:variance_gen_image_Fock_500_samples_theta_0.00000_no_errors} or~\ref{fig:variance_gen_image_Fock_500_samples_theta_1.57080_no_errors}) alone is sufficient to distinguish between different photon number states. 

This classification scheme has been extended to coherent states (Sec.~\ref{sec:CS}), in a context where it is known that only CS with different values of $\alpha$ are present. Here, computation of the mean photon number $|\alpha|^{2}$ (see Fig.~\ref{fig:CS_gen_samples_mean_photon_number_no_errors}) or $\langle \hat{x} \rangle$ in the case of real $\alpha$ suffices to classify the samples. A similar procedure has been carried out to classify a collection of single photon-added CS $|\alpha,1\rangle$ (Sec.~\ref{sec:PACS}). This is done by computing either the mean photon number or the quadrature variances. 

We now address the situation when different classes of states are part of a training set, as in the case of the two CS $|{\sf g}_{1}(\alpha)\alpha\rangle$ and $|\beta_{\rm opt}\rangle$, together with the $1$-PACS $|\alpha,1\rangle$. As has been pointed out earlier in the context of the experiment~\cite{Fadrny:2024}, either the mean photon number or $\langle \hat{x} \rangle$ can distinguish between the two CS. The numerical value of $\alpha$ itself can be obtained by computing either the mean photon number or the higher-order moments directly from the PDFs comprising the tomogram. However, in the regime where $|\alpha| \geqslant 1$, the variance needs to be computed to distinguish between the three states (since the $1$-PACS is squeezed along the $x$-quadrature while the CS is not; see Figs.~\ref{fig:variance_PACS_m1_g1_alpha} and~\ref{fig:variance_gen_image_PACS_500_samples_theta_0.00000_g1alpha_alpha_beta_opt_no_errors_panel}). It is therefore possible to consider machine learning for tomograms as a viable alternative to machine learning for reconstructed states, in this context. 

It is conventional to use fidelity as a quantifier of the difference between two states after reconstruction. Since, our procedure bypasses state reconstruction, we have proposed a Wasserstein distance-based confusion matrix to quantify the `closeness' between the generated and true PDFs, that comprise the tomograms. The details are elaborated in Appendix~\ref{sec:appendix_Wasserstein_based_confusion_matrix}. Figure~\ref{fig:avg_W1_confusion_matrix_Fock_rel_err_0.025} clearly demonstrates that our classification procedure successfully predicts the true state labels from the randomly generated samples. This procedure can be extended to other states also. 

\subsection{Nonclassicality from tomograms\label{sec:nonclassicality}} 
It is important to note that the equivalent of Wigner negativity can also be recognized directly from tomograms, using a nonclassicality indicator ${\rm NI}$~\cite{Rohith:2023}, based on the `nonclassical area' projected by the optical tomogram of the quantum state of light on the tomographic plane. 
\begin{figure}[!ht]
\centering
\includegraphics[width=0.45\textwidth]{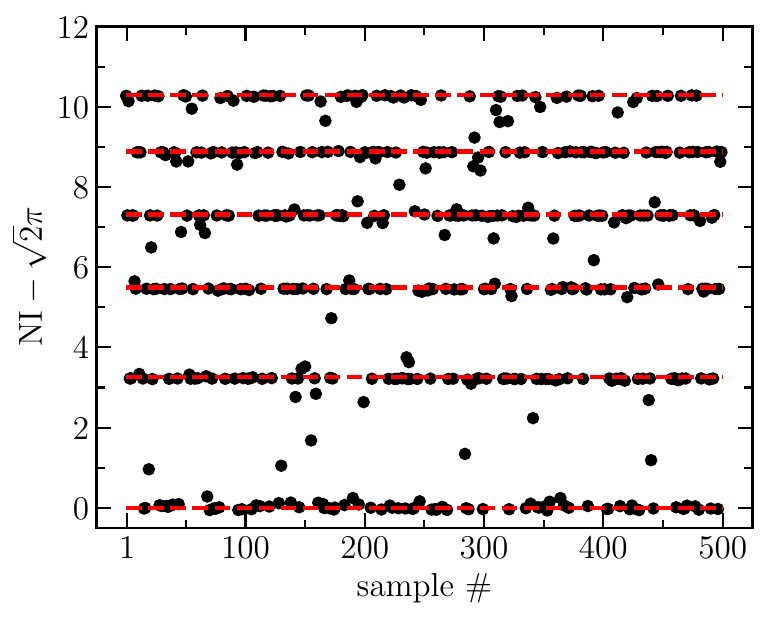}
\caption{Plot of ${\rm NI} - \sqrt{2}\pi$ versus generated samples (the same $500$ samples used in Fig.~\ref{fig:Fock_gen_samples_mean_photon_number_no_errors}) for the Fock states. Bottom to top $|n\rangle$: $n=0,1,2,\dots,5$. ${\rm NI}$ has been calculated using the quadrature variances $\Delta {\hat X}_{\theta}^{2}$ and Eq.~(\ref{eq:nonclassicality_indicator}). We have verified that this is $\theta$ independent for Fock states within $4\%$ tolerance. The red dashed lines are the theoretical values of ${\rm NI} - \sqrt{2}\pi$, and the black circles are the values computed from the generated tomograms. We see from the black circles that for most samples ${\rm NI} - \sqrt{2}\pi$ is either zero or close to zero if $n = 0$, and greater than zero only for $n > 0$.}
\label{fig:nonclassicality_indicator_Fock_no_errors}
\end{figure}

In terms of the quadrature variance $\Delta {\hat X}_{\theta}^{2}$ 
\begin{equation}
{\rm NI} = \int_{0}^{2\pi} d\theta \,\, \sqrt {\Delta {\hat X}_{\theta}^{2}} = \int_{-\pi}^{\pi} d\theta \,\, \sqrt {\Delta {\hat X}_{\theta}^{2}}, 
\label{eq:nonclassicality_indicator}
\end{equation}
and has a lower bound given by $\sqrt{2}\pi$. It can be readily seen by computing ${\rm NI}$ that the standard CS as well as $|0\rangle$ satisfy this lower bound. Now consider the Fock states $|n\rangle$, with variance $\Delta {\hat X}_{\theta}^{2} = (n+1/2)$ for every $\theta$. Equation~(\ref{eq:nonclassicality_indicator}) is an $n$-dependent quantity given by $\sqrt{2(2n+1)}\pi$. As shown in Fig.~\ref{fig:nonclassicality_indicator_Fock_no_errors}, in the case of Fock state tomograms generated with WGAN, (${\rm NI} - \sqrt2{\pi}$) is greater than zero only for $n > 0$, and increases with increase in $n$. This provides us with a method to quantify the nonclassicality of the state directly from its generated tomogram, without resorting to QST procedures. 

With the aid of a pre-computed glossary this approach can be readily extended to other sets of states. Whereas the standard procedure to classify quantum states into their respective categories involves both state reconstruction from the tomogram and an additional classifier neural network, our approach circumvents both these features. 

\section{Concluding remarks\label{sec:concluding_remarks}}
We have investigated the possibility of using a machine learning-based approach to characterize and classify quantum states of light directly from their optical tomograms,  without resorting to detailed state reconstruction. Our approach is based on the WGAN algorithm, in which two deep convolutional neural networks are trained simultaneously (on noise-free as well as noisy optical tomograms). The machine has been trained with tomograms corresponding to the photon-number states $|n\rangle$, the CS $|\alpha\rangle$ and the $1$-PACS $|\alpha,1\rangle$ for a range of values of $n$ and real $\alpha$. 

We emphasize that we have used WGAN only for generating tomograms which replicate the original input tomograms reasonably well. The mean photon number and the quadrature variances have been computed directly from the PDFs (in appropriate quadratures of both the input and generated tomograms), without involving WGAN, and without using an auxiliary network. 

In a recent experimental report, classical and quantum amplification have been investigated with reconstructed states of an amplified CS $|{\sf g}_{1}(\alpha)\alpha\rangle$, an optimal CS $|\beta_{\rm opt}\rangle$, and $|\alpha,1\rangle$ obtained by conditional addition of a photon. The range of values of $\alpha$ for which the fidelity computed between these states is close to unity, has been identified.  In the tomographic setting this would imply that the mean photon numbers computed from the generated tomograms of these three states should not merely be close to their theoretically expected values, but should also approach each other for values of $\alpha$ corresponding to high fidelity. This has been successfully demonstrated in our procedure. All the generated numerical values lie within $4\%$ of the expected theoretical values not only in the context of the experiment, but also when the generated Fock states,  the CS and the $1$-PACS were characterized after training the machine. We have verified that a tomogram with $128 \times 128$ pixels suffices to characterize a state. 

We have shown that the results obtained with three different choices of colormaps, namely, a sequential linear colormap, its nonlinear counterpart, and a nonlinear colormap with a regulator are not significantly different from each other. At no point in our analysis with colormaps have we used the fact that the state is known, as an important input. We emphasize, that the PDFs comprising the generated tomograms are all that are required for characterizing the state, independent of whether it is a known or unknown state. Hence, a linear colormap is a reasonably good initial choice, irrespective of whether we have any prior information about the quantum state or not. We have shown that carrying out computation with raw tomographic data (numerical arrays) lead to results which are only comparable to, and not better than those obtained using tomographic images. 

We have also demonstrated that quantum states can be classified directly from their optical tomograms by computing the appropriate moments and comparing them with a `look-up table'. 
This approach does not require any additional classifier network. We have demonstrated using two error models that the network is robust against noise. 

Despite the high overall accuracy that has been achieved, our analysis revealed that the distribution of the mean photon number and the variances about their theoretical values, exhibits a noticeable skewness or bias, particularly evident in regions corresponding to gradual color changes or transitions within the tomograms. This is an inherent characteristic originating from color-transitional effects within the generator's architecture, which can introduce subtle inconsistencies in the generated patterns. 

Our work opens up several promising avenues for further research. To address the observed bias and reduce generation errors, future work could explore methodologies specifically designed to mitigate these color transitional effects. A particularly promising direction is the application of local padding techniques~\cite{Abdellatif:2024} where locally padding in convolutional layers can help synthesize neighboring patch information to fill in padded regions, improving color consistency at the edges, etc. By sharing border features between generated patches, such attempts could have the potential to improve consistency at boundaries,  such as those at the tails of the PDFs of photon-added states, and could possibly lead to reduction in the observed errors. One could also test other techniques such as color correction GANs or line art integration to guide color placement and boundary definitions, or super resolution GAN~\cite{Ledig:2016} which can also help mitigate edge effects. Further improvements can also be sought through systematic hyperparameter optimization for the neural networks. In addition, future work could consider real experimental tomograms (as in Appendix~\ref{sec:appendix_real_expts}), which are inherently noisy, to exploit the denoising capability of the WGAN. The generator acts as a denoiser by removing artifacts from densities obtained using traditional reconstruction methods, such as MLE~\cite{Huang:2022}. It is also possible to extend our method to generate distributions not present in the training dataset, which is an inherent property of the WGAN algorithm, since the generator learns a continuous mapping from the distribution of the (finite) training tomograms~\cite{Ditria:2020, Koch:2022}. 

The tomographic approach could provide a readily implementable procedure to characterize unknown states of light with an optimal number of slices. It could be a viable alternative to detailed state reconstruction when the Hilbert space associated with the system is large, as in the case of  spin arrays, multipartite qubit systems and hybrid quantum systems. 

\begin{acknowledgments}
We acknowledge partial support through funds from Mphasis to the Center for Quantum Information, Communication, and Computing (CQuICC), Indian Institute of Technology Madras. VB and SL thank the Department of Physics, Indian Institute of Technology Madras, for infrastructural support. SP acknowledges discussion with Aryan Garg on certain aspects of the WGAN algorithm. 
\end{acknowledgments}

\appendix
\section{Numerical arrays as inputs and outputs of the network\label{sec:appendix_numerical_arrays}} 
\begin{figure}
\centering
\includegraphics[width=0.45\textwidth]{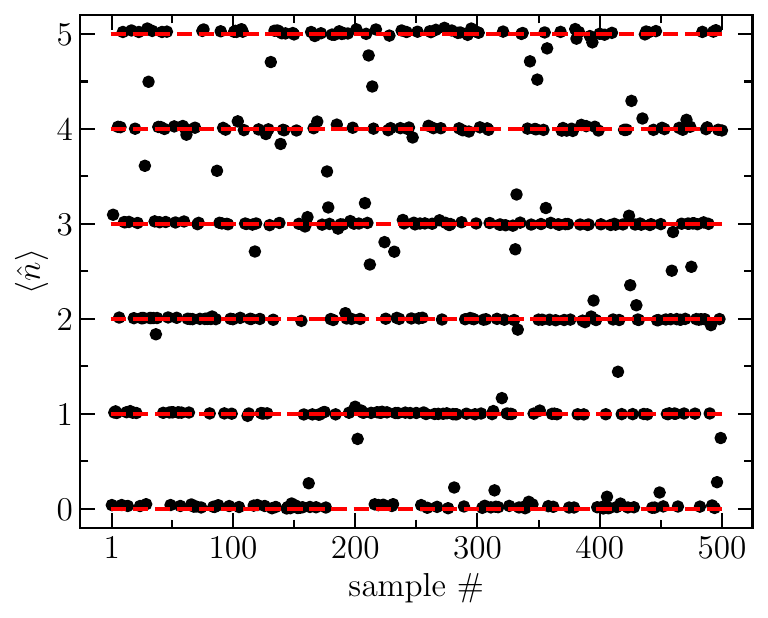}
\caption{The mean photon number $\langle \hat{n} \rangle$ (black circles) for the Fock states $|n\rangle~(n = 0, 1, \dots, 5)$ computed from $500$ randomly chosen generated tomograms. $\langle \hat{n} \rangle$ has been calculated from the PDFs along $\theta=0, \pi/3$ and $2\pi/3$, by setting $m=n=1$ in Eq.~(\ref{eq:mean_ph_num_Wunsche}). The red dashed lines are the theoretical values of $\langle \hat{n} \rangle$ corresponding to the Fock states considered. The computed value of $\langle \hat{n} \rangle$ is within $4\%$ error tolerance of the theoretical value for each generated state whose tomogram can be identified to clearly correspond to a particular photon number. Spurious samples (typically black circles midway between two consecutive horizontal red lines) do not fall under this category. These arise due to inherent issues in the generation process.}
\label{fig:Fock_gen_samples_mean_photon_number_numerical_arrays_no_errors}
\end{figure}

\begin{figure}
\centering
\includegraphics[width=0.45\textwidth]{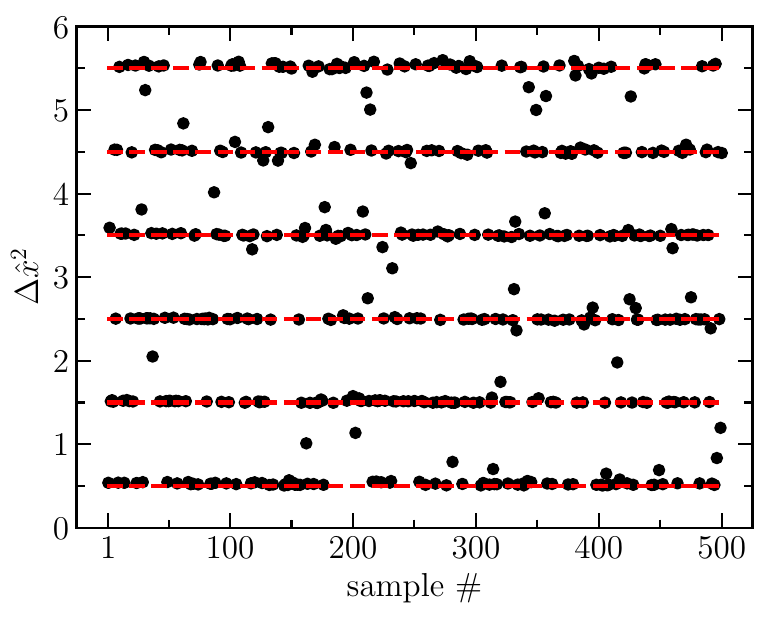}
\caption{The quadrature variance $\Delta \hat{x}^{2}$ (black circles) versus the sample number for $|n\rangle~(n=0,1,\dots,5)$ corresponding to $500$ randomly chosen generated samples, along the $x$-quadrature. This is evaluated using the PDFs for $\theta=0$. The horizontal red dashed lines are the theoretical values of $\Delta \hat{x}^{2}$ corresponding to the Fock states considered. Most of the generated samples are within $4\%$ error tolerance of the theoretical values. Spurious samples arise due to inherent issues in the generation process.}
\label{fig:variance_gen_image_Fock_500_samples_theta_0.00000_numerical_arrays_no_errors}
\end{figure}

For completeness, we report the results obtained using numerical arrays (raw tomographic data) as inputs and outputs of the network instead of images (RGB data). For demonstration purposes, we present the results in this case for Fock states (see Figs.~\ref{fig:Fock_gen_samples_mean_photon_number_numerical_arrays_no_errors} and~\ref{fig:variance_gen_image_Fock_500_samples_theta_0.00000_numerical_arrays_no_errors}). The network architecture (Tables~\ref{tab:generator_params} and~\ref{tab:discriminator_params}) was kept unchanged, with only the input (resp. output) channel dimension modified from $3$ to $1$ for the Discriminator (resp. Generator) network, as required. This facilitates comparison of the performance when using tomographic images with that of numerical arrays. Comparing Figs.~\ref{fig:Fock_gen_samples_mean_photon_number_numerical_arrays_no_errors} and~\ref{fig:variance_gen_image_Fock_500_samples_theta_0.00000_numerical_arrays_no_errors} with their counterparts (Figs.~\ref{fig:Fock_gen_samples_mean_photon_number_no_errors} and~\ref{fig:variance_gen_image_Fock_500_samples_theta_0.00000_no_errors}) obtained using tomographic images, it is evident that the two procedures lead to similar results. We have verified that this holds also in the case of the CS and the $1$-PACS. 


\section{Choice of error tolerance\label{sec:appendix_choice_of_err_tol}}
In what follows, we provide details of the procedure used to determine the appropriate relative error, and hence the resolution required for correctly classifying the states in our work. The classification, as mentioned earlier, relies on both the mean photon number and the quadrature variance. Our aim is to choose an error tolerance such that the {\it same} large subset of generated samples has both the mean photon number and quadrature variance within this error tolerance. This procedure is general enough to be used for other states as well. 

\begin{figure}[h!]
\centering
\includegraphics[width=0.45\textwidth]{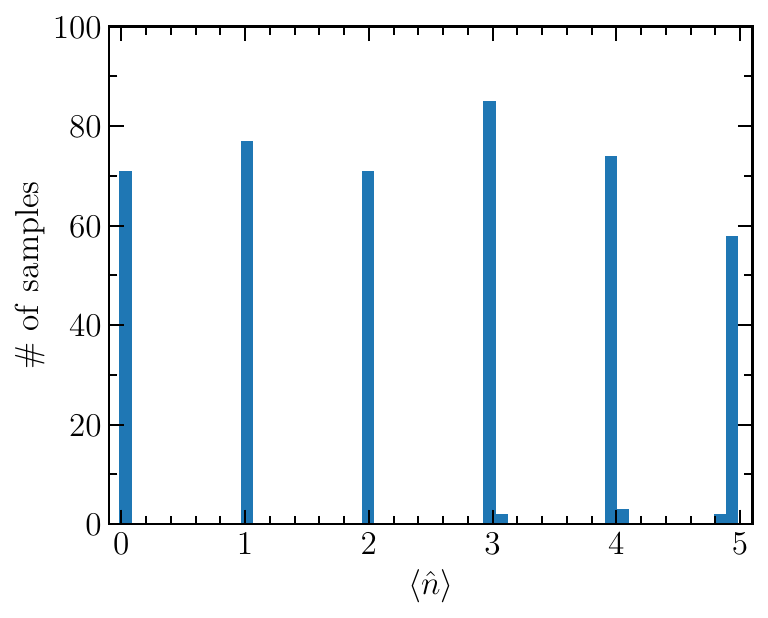}
\caption{Histograms of the $500$ randomly generated photon number states $|n\rangle \, (n=0.1,2,\dots,5)$. 
The relative error threshold is set to $2.5\%$ with respect to the theoretical expected values. 
It is evident that the majority of the samples ($443$ out of $500$) are correctly classified.}
\label{fig:histogram_all_generated_samples_Fock_rel_err_0.025}
\end{figure}
For illustrative purposes, consider the set of generated Fock state tomograms presented in Figs.~\ref{fig:Fock_gen_samples_mean_photon_number_no_errors} and~\ref{fig:variance_gen_image_Fock_500_samples_theta_0.00000_no_errors}. The mean photon number and the quadrature variance $\Delta {\hat x}^{2}$ are computed, say for a set of randomly chosen $500$ samples. If the tolerance is fixed at $2.5\%$, we see from the histogram in Fig.~\ref{fig:histogram_all_generated_samples_Fock_rel_err_0.025} that the majority of samples  can be correctly classified based on the mean photon number alone. This is corroborated in Table~\ref{tab:Fock_mean_photon_num_no_errors} (second column), where $443$ samples are seen to be correctly classified. Hence $\sim 90\%$ of the samples have their mean photon number within this chosen limit. However only $\sim 80\%$ of these samples lie within this error tolerance as far as the quadrature variance $\Delta {\hat x}^{2}$ is concerned (third column of Table~\ref{tab:Fock_quad_variance_x_no_errors} showing $414$ samples). However, if the tolerance is increased to $4\%$, then $\sim 90\%$ of the samples satisfy this limit with respect to both the mean photon number and the quadrature variance (Table~\ref{tab:Fock_mean_photon_num_no_errors}, column three and Table~\ref{tab:Fock_quad_variance_x_no_errors}, column four). This has been verified for other quadratures and other states as well. Hence the motivation for the choice of $4\%$ error tolerance in our results. 

\begin{table}[h!]
\begin{center}
\begin{tabular}{|c|c|c|}
\hline
Fock state,  & $\#$ of samples & $\#$ of samples \\
$|n\rangle$ & (relative error = $2.5\%$) & (relative error = $4\%$) \\ 
\hline \hline 
$|0\rangle$ & 71 & 75 \\
\hline 
$|1\rangle$ & 71 & 79 \\
\hline 
$|2\rangle$ & 71 & 74 \\
\hline 
$|3\rangle$ & 87 & 90 \\
\hline 
$|4\rangle$ & 77 & 79 \\
\hline 
$|5\rangle$ & 60 & 61 \\
\hline \hline 
Total & 443 & 458 \\
\hline 
\end{tabular}
\end{center}
\caption{The number of generated Fock states $|n\rangle$ classified into their respective $n$ values is shown for (a) $2.5\%$ error (second column), and (b) $4\%$ error (third column). The classification is carried out by evaluating the mean photon number $\langle \hat{n} \rangle$ of each generated sample and assigning the samples to the corresponding $n$ value. The samples considered here correspond to those shown in Figs.~\ref{fig:Fock_gen_samples_mean_photon_number_no_errors} and~\ref{fig:variance_gen_image_Fock_500_samples_theta_0.00000_no_errors}. It is evident that a large fraction ($> 90\%$) of the generated samples are correctly categorized under the chosen relative error, with larger relative errors resulting in a greater number of samples being included.}
\label{tab:Fock_mean_photon_num_no_errors}
\end{table}
\begin{table}[h!]
\begin{center}
\begin{tabular}{|c|c|c|c|}
\hline
Fock state, & $\Delta \hat{x}^{2}$  & $\#$ of samples & $\#$ of samples \\ 
$|n\rangle$ & & (relative error = $2.5\%$) & (relative error = $4\%$) \\ 
\hline \hline 
$|0\rangle$ & $1/2$ & 45 & 65 \\
\hline 
$|1\rangle$ & $3/2$ & 75 & 78 \\
\hline 
$|2\rangle$ & $5/2$ & 71 & 74 \\
\hline 
$|3\rangle$ & $7/2$ & 86 & 89 \\
\hline 
$|4\rangle$ & $9/2$ & 78 & 81 \\
\hline 
$|5\rangle$ & $11/2$ & 59 & 61 \\
\hline \hline 
Total & & 414 & 448 \\
\hline 
\end{tabular}
\end{center}
\caption{The number of generated Fock states $|n\rangle$ classified into their respective $\Delta \hat{x}^{2}$ values is shown for (a) $2.5\%$ error (second column), and (b) $4\%$ error (third column). The classification is carried out by calculating $\Delta \hat{x}^{2}$ for the same set of samples corresponding to each $n$ value in Table~\ref{tab:Fock_mean_photon_num_no_errors}. It is clear that setting a relative error of $4\%$ (with reference to the theoretical value of $\Delta \hat{x}^{2}$) results in approximately the same set of samples which were correctly identified in terms of the mean photon number.}
\label{tab:Fock_quad_variance_x_no_errors}
\end{table}

\section{Noisy optical tomograms\label{sec:appendix_noise_models}}
\begin{figure}[!ht]
\centering
\includegraphics[width=0.45\textwidth]{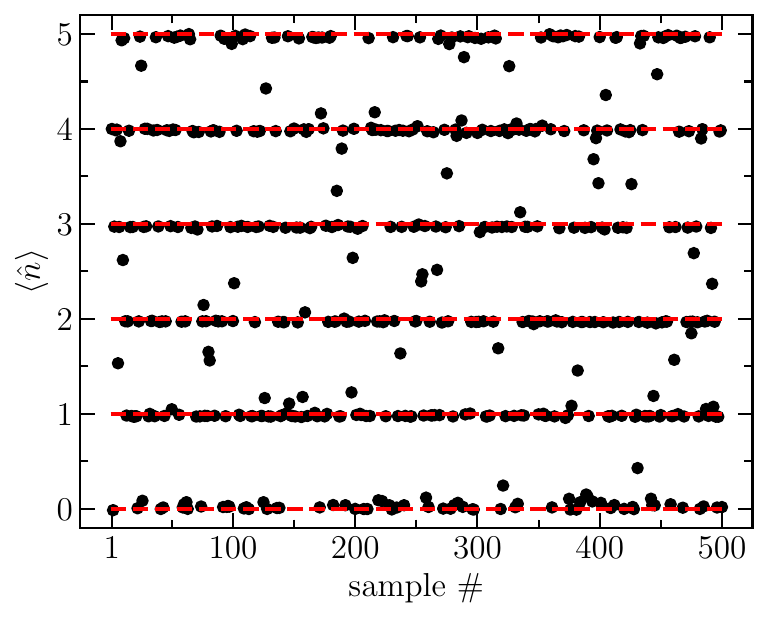}
\caption{The mean photon number $\langle \hat{n} \rangle$ for the photon number states $|n\rangle$ ($n=0,1,\dots, 5$) corresponding to $500$ randomly chosen generated samples. $\langle \hat{n} \rangle$ has been computed using the procedure mentioned in Sec.~\ref{sec:wgan_network_arch_dataset_prep}. The red horizontal dashed lines are the theoretical (expected) values of $\langle \hat{n} \rangle$ corresponding to the Fock states considered. The training dataset consists of optical tomograms with noise introduced through model (b) setting $\epsilon=0.25$. By comparing with Fig.~\ref{fig:Fock_gen_samples_mean_photon_number_no_errors} we see that the introduction of errors in the training dataset has not significantly altered the results.}
\label{fig:Fock_gen_samples_mean_photon_number_error_model_b_eps_0.25}
\end{figure}
\begin{figure}[!ht]
\centering
\includegraphics[width=0.45\textwidth]{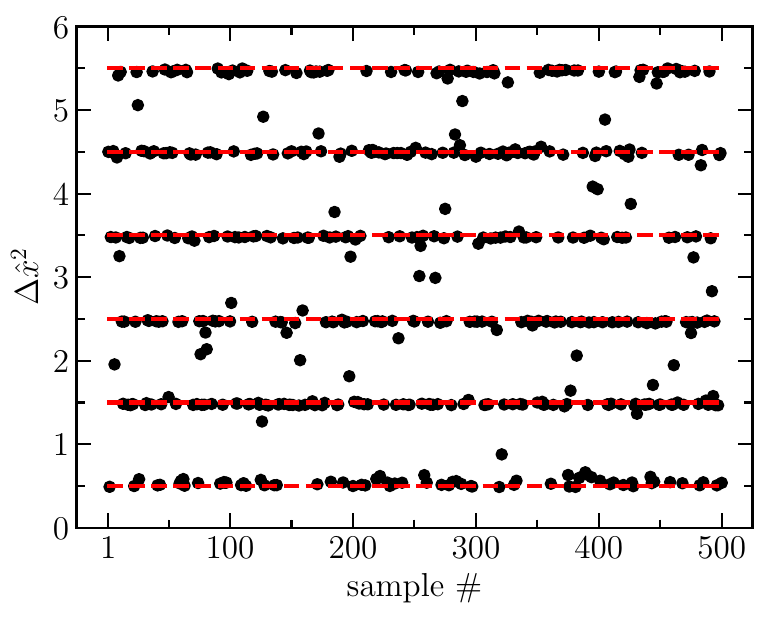}
\caption{The quadrature variance $\Delta \hat{x}^{2}$ for the photon number states $|n\rangle~(n=0,1,\dots,5)$ corresponding to $500$ randomly chosen generated samples, along the $x$-quadrature. The horizontal red dashed lines are the theoretical (expected) values of $\Delta \hat{x}^{2}$ corresponding to the Fock states considered. This is evaluated using the reconstructed PDFs along $\theta=0$. The training dataset consists of optical tomograms with noise introduced through model (b) setting $\epsilon=0.25$. By comparing with Fig.~\ref{fig:variance_gen_image_Fock_500_samples_theta_0.00000_no_errors} we see that the introduction of errors in the training dataset has not significantly altered the results.}
\label{fig:variance_gen_image_Fock_500_samples_theta_0.00000_error_model_b_eps_0.25}
\end{figure}
The effects of the two error models described in Sec.~\ref{sec:wgan_network_arch_dataset_prep}, on the mean photon number and $\Delta \hat{x}^{2}$ corresponding to the photon number states $|n\rangle$ ($n=0,1,2,\dotsc,5$) are outlined below. Post training, $500$ randomly generated samples were chosen, and $\langle \hat{n} \rangle$ and the quadrature variances $\Delta \hat{X}_{\theta}^{2}$ ($\theta=0, \pi/3, \pi/2, 2\pi/3$) were computed. 
We have verified that both $\langle \hat{n} \rangle$ and $\Delta \hat{x}^{2}$ are within reasonable error tolerances, for both error models and for both values of $\epsilon$ considered. This indicates robustness to noise. This is consistent with results on noise in neural networks reported in the literature~\cite{Nazare:2018}. These conclusions hold for quadrature variances for $\theta=\pi/3, \pi/2, 2\pi/3$ as well. As proof of concept $\langle \hat{n} \rangle$ and $\Delta \hat{x}^{2}$ are presented for model (b) setting $\epsilon = 0.25$ in Figs.~\ref{fig:Fock_gen_samples_mean_photon_number_error_model_b_eps_0.25} and ~\ref{fig:variance_gen_image_Fock_500_samples_theta_0.00000_error_model_b_eps_0.25} respectively. 

\section{Training on experimental data\label{sec:appendix_real_expts}}
As mentioned in the text, we have used the PDFs (tomograms) from two experiments on single-photon production and characterization~\cite{Hsieh:2024, Zavatta:2004b}, as inputs for training. 
Out of the several output tomograms that were generated after training, $500$ samples were randomly selected. 

\begin{figure}
\centering
\includegraphics[width=0.45\textwidth]{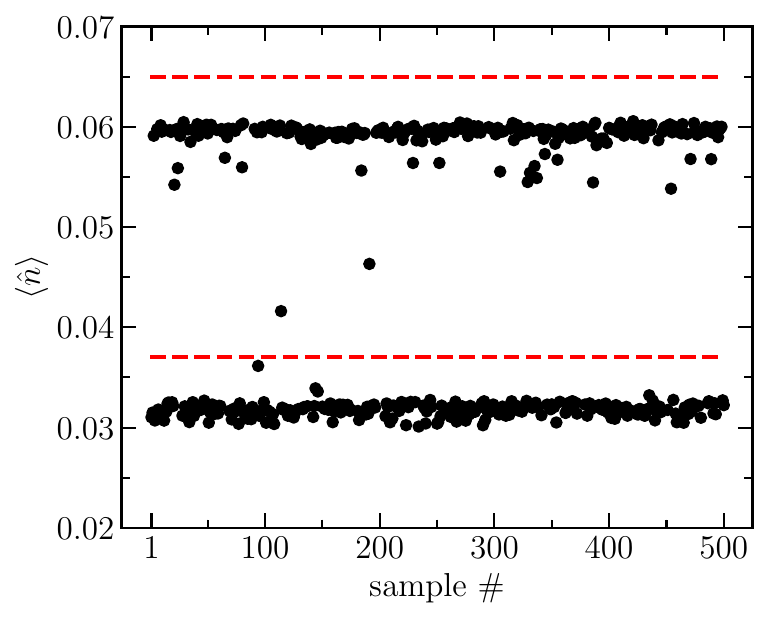}
\caption{The mean photon number $\langle \hat{n} \rangle$ (black circles) corresponding to the experimental report (Eq.~(\ref{eq:expt_Fock_1_PDF_model})) computed from $500$ randomly chosen generated tomograms. $\langle \hat{n} \rangle$ has been calculated from the PDFs along $\theta=0, \pi/3$ and $2\pi/3$, by setting $m=n=1$ in Eq.~(\ref{eq:mean_ph_num_Wunsche}). The red dashed lines are the theoretical values of $\langle \hat{n} \rangle = 0.037$ and $0.065$, corresponding to $\eta=0.574$ and $0.631$, respectively.}
\label{fig:Fock_gen_samples_mean_photon_number_Fock_1_expt}
\end{figure}

\begin{figure}
\centering
\includegraphics[width=0.45\textwidth]{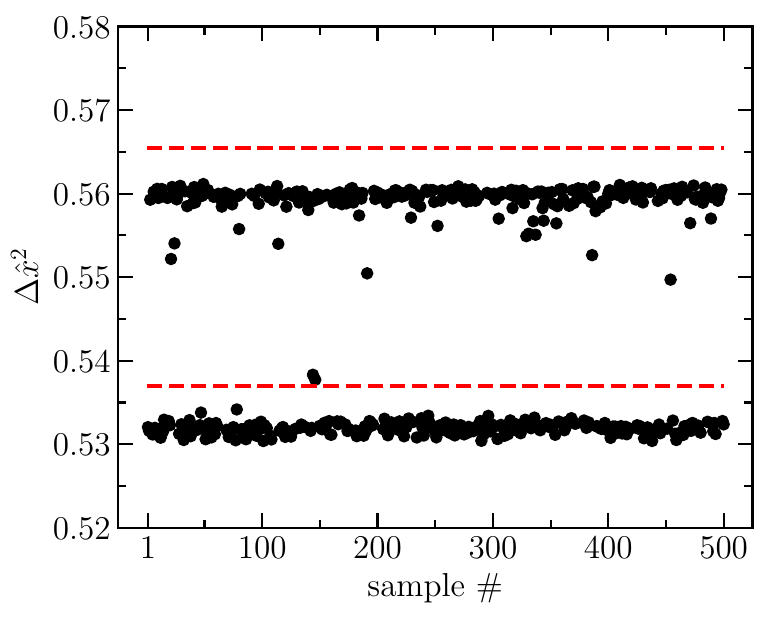}
\caption{The quadrature variance $\Delta \hat{x}^{2}$ (black circles) corresponding to the experimental report (Eq.~(\ref{eq:expt_Fock_1_PDF_model})) for each of the generated samples, along the $x$-quadrature. This is evaluated using the reconstructed PDFs along $\theta=0$. The red dashed lines represent the theoretically computed values $0.537$ and $0.566$, corresponding to $\eta=0.574$ and $0.631$, respectively.}
\label{fig:variance_gen_image_Fock_500_samples_theta_0.00000_Fock_1_expt}
\end{figure}

Figures~\ref{fig:Fock_gen_samples_mean_photon_number_Fock_1_expt} and~\ref{fig:variance_gen_image_Fock_500_samples_theta_0.00000_Fock_1_expt} are respectively the mean photon number $\langle \hat{n} \rangle$ and the quadrature variance $\Delta {\hat x}^{2}$ plotted as functions of the generated sample numbers. Each black circle corresponds to a randomly generated sample. The red horizontal lines in Fig.~\ref{fig:Fock_gen_samples_mean_photon_number_Fock_1_expt} indicate the theoretically expected mean photon numbers $\langle {\hat n} \rangle = 0.037$ and $0.065$, for $\eta = 0.574$ and $0.631$, respectively. Similarly in Fig.~\ref{fig:variance_gen_image_Fock_500_samples_theta_0.00000_Fock_1_expt}, the red horizontal lines indicate the theoretical quadrature variance $\Delta {\hat x}^{2} = 0.537$ and $0.566$, for $\eta = 0.574$ and $0.631$, respectively. It is evident from both plots that states produced in two distinct realistic experiments can be reliably classified and distinguished from each other using our procedure. 

\section{Wasserstein distance-based confusion matrix\label{sec:appendix_Wasserstein_based_confusion_matrix}}
\begin{figure}[t]
\centering
\includegraphics[width=0.45\textwidth]{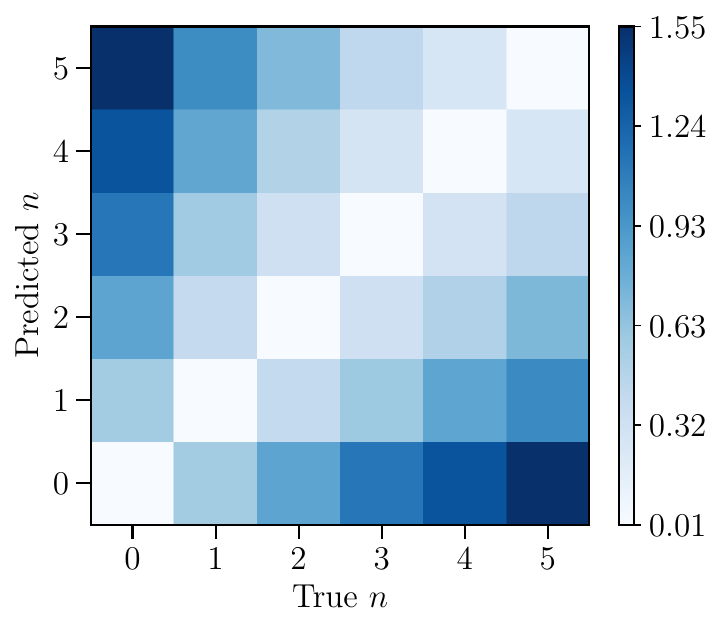}
\caption{The Wasserstein distance-based confusion matrix for the generated photon number states $|n\rangle \, (n=0,1,2,\dots,5)$. 
First, a relative error threshold of $2.5\%$ about the expected (true) value of both the mean photon number and the quadrature variance was applied to the corresponding quantities obtained from the generated samples for each Fock state $|n\rangle$. The samples (labeled by their $n$ value) that satisfied this criterion were subsequently selected for calculating the confusion matrix. For every generated sample belonging to the Fock state $|n\rangle$, the Wasserstein distance between the generated PDF and the theoretical PDF, along the $x$-quadrature, was computed. The average of all these distances was finally used to compute the confusion matrix. It is evident that when the predicted value of $n$ from the generated sample is equal to the true $n$, $W_{1} \sim 0$ (diagonal entries).}
\label{fig:avg_W1_confusion_matrix_Fock_rel_err_0.025}
\end{figure}

To quantify the performance of our classification procedure, we have now proposed and evaluated a Wasserstein distance ($W_{1}$) based confusion matrix (Fig.~\ref{fig:avg_W1_confusion_matrix_Fock_rel_err_0.025}) for the generated Fock state samples in Fig.~\ref{fig:Fock_gen_samples_mean_photon_number_no_errors}. It is based on the fact that when states are close, i.e., the corresponding fidelity $\sim 1$, the $W_{1}$ (in appropriate quadrature) between them approaches zero. To evaluate this matrix, we first set a relative error threshold of $2.5\%$ around the expected (true) values of both the mean photon number and the quadrature variance. We have considered the corresponding quantities computed from the generated samples for each Fock state $|n\rangle\,(n=0,1,2,\dots5)$. Samples (labeled by their respective $n$ values) that satisfied this error criterion were then retained for the construction of the confusion matrix. For each retained sample associated with the Fock state $|n\rangle$, the Wasserstein distance between the generated and theoretical probability density functions along the $x$-quadrature was evaluated. The confusion matrix was constructed using the average of these distances. As expected, the classification performs well, with near-zero diagonal entries, and non-zero off-diagonal entries. A similar exercise can be carried out for the other states as well. 

\section{Variance in the $p$-quadrature\label{sec:appendix_variances_p_quadratures}}
\begin{figure}[h]
\centering
\includegraphics[width=0.45\textwidth]{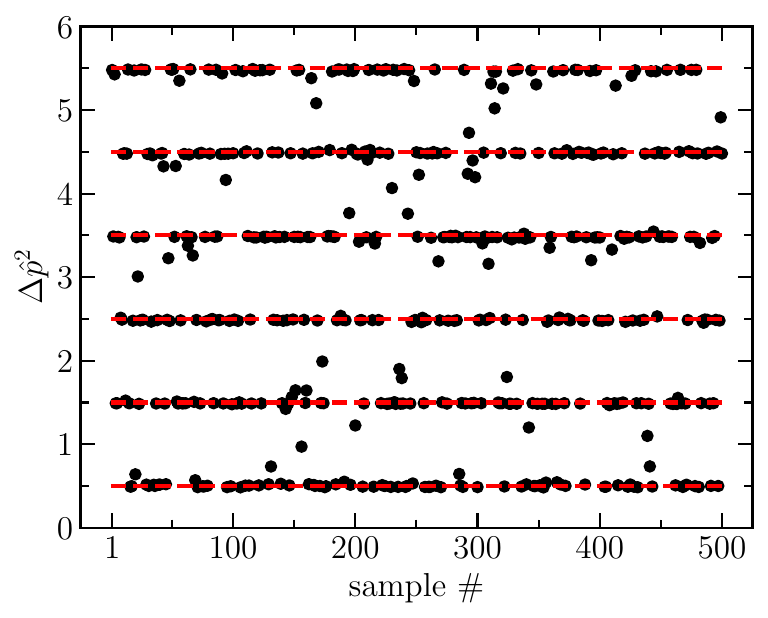}
\caption{The quadrature variance $\Delta \hat{p}^{2}$ (black circles) versus the sample number for $|n\rangle~(n=0,1,\dots,5)$ corresponding to $500$ randomly chosen generated samples, along the $p$-quadrature. This is evaluated using the reconstructed PDFs for $\theta=\pi/2$. The horizontal red dashed lines are the theoretical values of $\Delta \hat{p}^{2}$ corresponding to the Fock states considered. Most of the generated samples are within $4\%$ error tolerance of the theoretical values. Spurious samples arise due to inherent issues in the generation process.}
\label{fig:variance_gen_image_Fock_500_samples_theta_1.57080_no_errors}
\end{figure}
\begin{figure}[!h]
\centering
\includegraphics[width=0.45\textwidth]{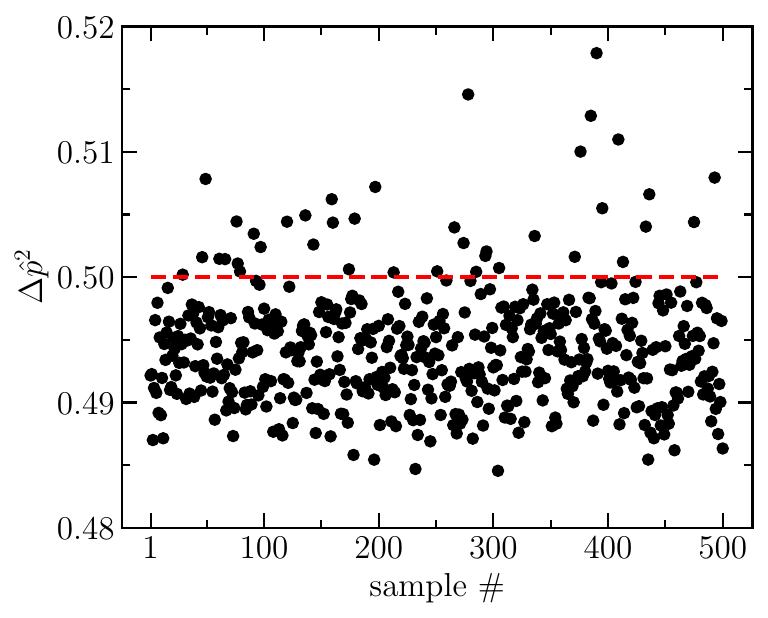}
\caption{The quadrature variance $\Delta \hat{p}^{2}$ (black circles) for the CS $|\alpha\rangle$ corresponding to each of the generated samples, along the $p$-quadrature. This is evaluated using the reconstructed PDFs along $\theta=\pi/2$. The red dashed line represents the theoretical value $0.5$, independent of the value of $\alpha$. As in the case of Fock states, most of the black circles (generated values of the variance) lie within $4\%$ of $0.5$. The skewness about the horizontal red dashed line is discussed in Sec.~\ref{sec:nonlinear_cmap_CS}.}
\label{fig:variance_gen_image_CS_500_samples_theta_1.57080_no_errors}
\end{figure}
\begin{figure}[!h]
\centering
\includegraphics[width=0.45\textwidth]{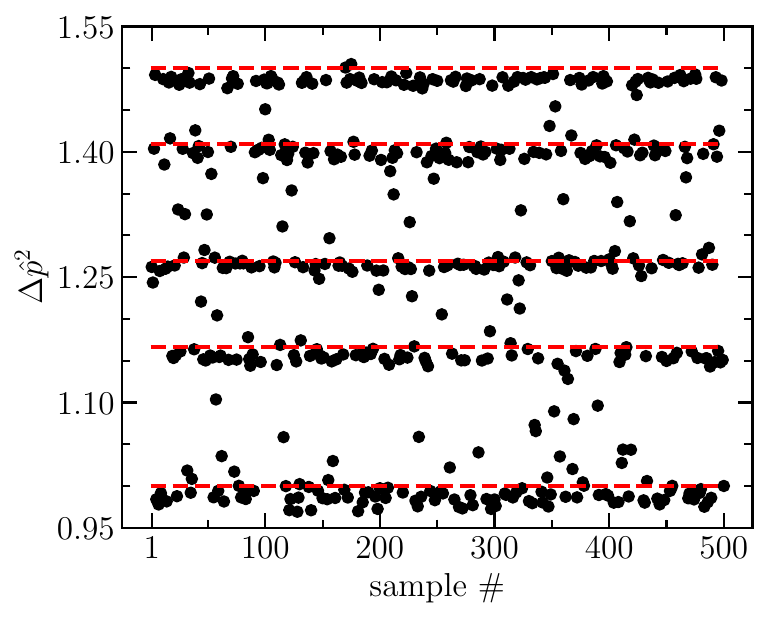}
\caption{The quadrature variance $\Delta \hat{p}^{2}$ (black circles) for $|\alpha,1\rangle$ corresponding to each of the generated samples, along the $p$-quadrature. This is evaluated using the reconstructed PDFs along $\theta=\pi/2$. The red dashed lines represent the theoretically computed values $1.50, 1.41, 1.27, 1.17$, and $1$, corresponding to $\alpha^{2}=0, 0.1, 0.3, 0.5$ and $1$, respectively.}
\label{fig:variance_gen_image_PACS_500_samples_theta_1.57080_no_errors}
\end{figure}
The quadrature variance $\Delta \hat{p}^{2}$ corresponding to $\theta=\pi/2$ has been evaluated using the reconstructed PDFs of the Fock states $|n\rangle$ (Fig.~\ref{fig:variance_gen_image_Fock_500_samples_theta_1.57080_no_errors}), the CS $|\alpha\rangle$ (Fig.~\ref{fig:variance_gen_image_CS_500_samples_theta_1.57080_no_errors}) and the $1$-PACS $|\alpha,1\rangle$ (Fig.~\ref{fig:variance_gen_image_PACS_500_samples_theta_1.57080_no_errors}). The variance was computed for the $500$ randomly generated samples with which $\Delta \hat{x}^{2}$ and $\langle \hat{n} \rangle$ were obtained. Results using the sequential linear colormap, and the nonlinear colormap (with regulator) agree with each other. 

The red horizontal dashed lines in Figs.~\ref{fig:variance_gen_image_Fock_500_samples_theta_1.57080_no_errors},~\ref{fig:variance_gen_image_CS_500_samples_theta_1.57080_no_errors} and~\ref{fig:variance_gen_image_PACS_500_samples_theta_1.57080_no_errors} denote the theoretical values of $\Delta \hat{p}^{2}$ (equal to $n+1/2$ for the Fock states, $0.5$ for the CS independent of the value of $\alpha$, and $1.5, 1.41, 1.27, 1.17$ and $1$, for $\alpha=0, \sqrt{0.1}, \sqrt{0.3}, \sqrt{0.5}$ and $1$ respectively, in the case of $|\alpha,1\rangle$). 

Each black circle is the variance in the $p$-quadrature computed from one of the corresponding output tomograms. As in the case of $\Delta \hat{x}^{2}$ obtained earlier, the black circles are within $4\%$ of the theoretical values. The skewness observed in the distribution of black circles about the theoretical value of $\Delta \hat{x}^{2}$ in Figs.~\ref{fig:variance_gen_image_CS_500_samples_theta_0.00000_no_errors} and~\ref{fig:variance_CS_alpha_0.00000_0.54772_theta_0.00000_nlin_cmap_panel_no_errors_panel} in the case of the CS can also be seen in Fig.~\ref{fig:variance_gen_image_CS_500_samples_theta_1.57080_no_errors}, corresponding to $\Delta \hat{p}^{2}$. As pointed out earlier, this is probably a consequence of the GAN architecture. 


\newpage 
\bibliography{references} 

\end{document}